\documentclass[twocolumn]{bmcart} 
\usepackage[utf8]{inputenc}
\usepackage{amsmath}
\usepackage{hyperref}
\usepackage{graphicx}

\startlocaldefs
\endlocaldefs

\begin{document}

\begin{frontmatter}

\begin{fmbox}
\dochead{Research}

\title{Modelling railway delay propagation as diffusion-like spreading}

\author[
   addressref={aff1,aff2},                   
   corref={aff1},                       
   email={m.m.dekker@uu.nl}   
]{\inits{MMD}\fnm{Mark M.} \snm{Dekker}}
\author[
   addressref={aff3},
]{\inits{AM}\fnm{Alexey N.} \snm{Medvedev}}
\author[
   addressref={aff4},
]{\inits{JR}\fnm{Jan} \snm{Rombouts}}
\author[
   addressref={aff5},
]{\inits{GS}\fnm{Grzegorz} \snm{Siudem}}
\author[
   addressref={aff6,aff7},
]{\inits{LT}\fnm{Liubov} \snm{Tupikina}}
\address[id=aff1]{%
  \orgname{m.m.dekker@uu.nl, Department of Information and Computing Sciences, Utrecht University}, 
  \city{Utrecht},
  \cny{The Netherlands}
}
\address[id=aff2]{%
  \orgname{Centre for Complex Systems Studies, Utrecht University},
  \city{Utrecht},
  \cny{The Netherlands}
}
\address[id=aff3]{%
  \orgname{alexey.medvedev@unamur.be, ICTEAM, Université catholique de Louvain},
  \city{Louvain-la-Neuve},
  \cny{Belgium}
}
\address[id=aff4]{%
  \orgname{jan.rombouts@kuleuven.be, Laboratory of Dynamics in Biological Systems, KU  Leuven},
  \city{Leuven},
  \cny{Belgium}
}
\address[id=aff5]{%
  \orgname{grzegorz.siudem@pw.edu.pl, Warsaw University of Technology, Faculty of Physics},
  \city{Warsaw},
  \cny{Poland}
}
\address[id=aff6]{%
  \orgname{liubov.tupikina@cri-paris.org, Center for Research and Interdisciplinarity (CRI), Universit\'e de Paris},
  \city{Paris},
  \cny{France}
}
\address[id=aff7]{%
  \orgname{Nokia Bell labs},
  \city{Paris},
  \cny{France}
}
\begin{artnotes}
\end{artnotes}

\begin{abstractbox}
\begin{abstract}
Railway systems form an important means of transport across the world. However, congestions or disruptions may significantly decrease these systems' efficiencies, making predicting and understanding the resulting train delays a priority for railway organisations. Delays are studied in a wide variety of models, which usually simulate trains as discrete agents carrying delays. In contrast, in this paper, we define a novel model for studying delays, where they spread across the railway network via a diffusion-like process. This type of modelling has various advantages such as quick computation and ease of applying various statistical tools like spectral methods, but it also comes with limitations related to the directional and discrete nature of delays and the trains carrying them. We apply the model to the Belgian railways and study its performance in simulating the delay propagation in severely disrupted railway situations. In particular, we discuss the role of spatial aggregation by proposing to cluster the Belgian railway system into sets of stations and adapt the model accordingly. We find that such aggregation significantly increases the model's performance. For some particular situations, a non-trivial optimal level of spatial resolution is found on which the model performs best. Our results show the potential of this type of delay modelling to understand large-scale properties of railway systems.
\end{abstract}


\end{abstractbox}
\end{fmbox}

\end{frontmatter}

\section{Introduction}
\label{sec:1_intro}

Railway systems are of vital importance for transporting passengers and goods. The trains in these systems travel via predefined schedules that allow for highly efficient utilization of the routes and tracks. Temporal deviations from such scheduled operations are commonplace. They take the form of \textit{delays} and decrease the system's efficiency. Small delays are often absorbed by built-in buffers and therefore do not have effects on larger scales \cite{Zieger2018, dekker2021plos}. However, from time to time, logistic disruptions --- often caused by external factors like weather --- lead to congestion or even a large-scale stand-still, with detrimental costs to society and economy \cite{Ludvigsen2014, Tsuchiya2007, buchel2020, dekker2018}.

The above shows the importance of better understanding of delay propagation and its prediction. A large variety of delay propagation models exists, and the choice of the approach depends on a number of questions related to, among other factors, the spatial focus, availability of data and the delay severity. For example, when aiming to accurately predict the delay in a geographically confined area, there are high-performing statistical models \cite{kecman2015, li2016}. However, such statistics generally only work accurately in circumstances where delay is not too severe --- as per definition these highly delayed scenarios are exceptional. Also, when upscaling to larger areas, long-range interactions and associated correlations come into play which may be difficult to account for when using average statistics. Larger scales and more highly delayed scenarios are therefore often analyzed with machine learning or big-data approaches \cite{dekker2019, Oneto2018}, but at the cost of understanding cause-and-effect or fine spatial resolution. Alternatives to such purely data-driven methods can be found in models where mechanisms of delay propagation are explicitly implemented. For example, Monechi et al. analyzed the German and Italian railways and found a set of `laws' that drive the spreading of delays \cite{monechi2018}, analogous to epidemic spreading models. 
Of course, the infrastructure networks underlying the dynamical processes in any of the mentioned models play an important constraining role. However, this information is already embedded in the schedules and therefore less discussed in the context of delay simulations. However, the role of railway network topology is addressed by various scholars in relation to resilience properties \cite{sen2003, erath2009, bhatia2015}. Most models are based on the schedules of the railway system, commonly using trains as agents that have the potential to carry delays. The perspective of delays as a properties of discrete trains or events can be found in many analytical models \cite{goverde2010, Gambardella2002, Buker2012, Harrod2019, dekker2021plos}, using either deterministic or stochastic techniques to derive future delays from past information. Because of the abundance of this perspective in existing delay propagation models, we refer to the view of delays as properties of discrete trains or events as the `traditional view'. In contrast, one could also view delays as variables associated not to trains, but to the nodes (stations) and edges of the railway network, which stay in the same position. How delay spreads between these nodes does not have to be described in terms of discrete trains and events, but instead a description may rely solely on general (or even system-wide) quantities such as the network topology and schedule. One can make the analogy of fluid dynamics: while traditionally, delays are treated as Lagrangian particles (i.e., following the trains as the fluid carrying the particles), we propose to treat delays from an Eulerian point of view (i.e., determining incoming and outgoing delays in a fixed spatial frame). This is also discussed in \cite{Dekker2019ifac}. This is the basis of the model proposed in this paper.

The traditional view of delays as discrete quantities of explicitly modelled trains or events is useful because it allows for tracking expected routes of delays along the train's trajectories explicitly. In other words, given that you know that delay is in the system at location $A$, it is unlikely to spread in all possible directions from $A$, but more likely to follow a particular direction that is dependent on which trains are exactly affected. One only knows this direction if discrete train units (and their trajectories) are explicitly included in the model. But there are also disadvantages of such models. One limitation is that many such models rely on many statistics in addition to mere schedule information. For example, if trains $A$, $B$ and $C$ are simulated explicitly, the interactions of all their events and relative magnitudes of their delays have an impact on each other's delays. These relations need to be well studied using for example neural networks \cite{Oneto2017, Oneto2018} or probability updating \cite{Corman2018, berger2011}. Another limiting consideration of treating delays as discrete quantities is the spatial scale. In confined systems, the mechanisms of delay propagation and their parameters can be well-defined, as in \cite{kecman2015}. Defining all such interactions on a country-wide scale is generally more more complex, due to potential long-range correlations.

In this paper we propose to treat delays not as bound by discrete trains or events, but rather as continuously spreading across the infrastructure network. The spreading between nodes of the network is weighted by properties of the system. The intuition behind these models is that on average --- in a `mean-field approximation' --- these parameters drive the overall direction of delay propagation. We refer to this way of treating delay propagation as `diffusion-like spreading'. Small-scale accuracy is traded for larger scale accuracy: when looking at a micro-scale or individual trains, we expect this non-traditional way of dealing with delays to be less accurate than more detailed models, but on a large scale, we expect the performance of such a model to increase. As is shown in section 2, the model contains only simple schedule information (e.g., train frequencies and travel times) rather than complicated statistics, and all model information is embedded in a single matrix, which makes analysis of the system's properties easy. The mentioned reasons motivate us to write this paper on delay propagation as a diffusion-like spreading mechanism. We apply our proposed model to the Belgian railways as a case study to discuss when and how it is advantageous to use such models.

An important aspect of delay propagation in general is the spatial scale and resolution of the analysis. High resolution (`micro-scale') modelling allows for explicit simulation of infrastructure capacity issues, the role of speed gradients or the identification of station-specific properties, for example. Low resolution, but large-scale (`macro-scale') modelling captures the impact of long-range interactions related to resource allocation \cite{dekker2021plos}, the impact of long train lines \cite{dekker2019} or other system-wide properties. Many models lie between these extremes. Diffusion-like models should typically be regarded as having a lower resolution but working well on a larger scale, because of the earlier mentioned trade of small-scale accuracy for larger-scale accuracy. Spatial resolution is often expressed by treating railway infrastructure as a network, consisting of nodes (geographical locations) and edges (connections between them). At the highest spatial resolution, the nodes are certain control points in stations and tracks, where train activities are logged \cite{dekker2021plos}. More commonly is a slightly aggregated version of this, namely the more coarse passenger stations \cite{bhatia2015, goverde2010, monechi2018}. Lower resolutions are obtained when constructing regions that correspond to groups of stations --- so-called `clusters', on which we elaborate later. Larger geographical areas in lower resolutions combine existing delays from higher resolutions and are treated as one unit. Choosing the correct level of spatial aggregation is an important consideration to make when assessing the viability of the diffusion-like model.

When discussing spatial aggregation, it is important to define how higher levels of aggregation are derived from the lower ones. In particular: how do we join stations and tracks together into larger and coarser regions? A large amount of complex network literature is devoted to this question of \textit{clustering}, and clustering methods come in many forms in various applications \cite{Fortunato2016}. For example, graph or connectivity-based methods emphasize how connections and topology lead to a natural aggregation of nodes into larger groups. This can be quantified by the so-called modularity of the partition, first proposed by Newman \cite{Newman2006}. Various clustering algorithms based on modularity optimisation exist, such as the Louvain method \cite{Blondel2008}. 

Spectral clustering focuses on properties of the eigenspace of the Laplacian or model-relevant matrices. A third common method for clustering any --- also non-networked --- data is K-means \cite{Steinhaus1956, MacQueen1967}, which defines centroids and groups nodes based on their respective distances to these cluster centroids (also known as Voronoi iteration \cite{Lloyd1982}), given a definition of `distance' between nodes. This method has been used in the context of transportation before, albeit mostly to characterise statistical space (rather than actual stations and physical space) \cite{Kadir2018, Cerreto2018}. An important aspect of K-means, in contrast to for example the Louvain method, is that it requires the specification of the number of desired clusters ($K$) up front, which can be both advantageous and disadvantageous. However, the freedom of choosing $K$ turns out to be useful when analysing our diffusion-like delay model. This, together with the fact that K-means is a well known and commonly used method, motivates us to use K-means with geographical distance to cluster the stations in our paper. By choosing the number of clusters, we vary the spatial aggregation level. We will compare the performance of the diffusion-like model on each of these levels.

In summary, the aim of this paper is to discuss the usefulness of treating delay propagation as a diffusion-like spreading mechanism. We propose a model doing so in section 2. We apply the model to the example case of the Belgian railways and discuss the data and methodology for this in section 3. Section 4 discusses the results of a toy model, its performance on different types of disrupted situations, and the overall performance of the model. Here we discuss in what cases the diffusion-like aspect of the model is beneficial and what we can learn about the Belgian railways using this framework. We end with a summary and several conclusive remarks in section 5.

\section{Model}
\label{sec:2_model}

In this section we introduce our diffusion-like model. We start by defining the delay variable and set up the equations that describe its evolution over time. We continue by discussing how this model can be generalized to any spatial scale. For a detailed derivation of the model, see Appendix~A. Tab.~\ref{tab:paramdesc} summarizes the variables and parameters of the model. 

\subsection{General concepts}
\label{sec:modelsetup}

The main idea behind the model is to define the delay on fixed locations, and to describe the evolution of this delay distribution over time using macroscopic parameters such as train frequencies and travel times. While delays are inherent attributes of trains (i.e. agents), we aggregate the delays on passenger stations (i.e. nodes), as the impact of disruptions can mostly be felt at level of stations rather than being a problem of individual trains. This aggregation of delays onto stations means that we lose some of the finer details on which delays belong to which train. However, it will allow us to use tools for studying dynamical processes on networks: a delay is associated to each node, and its evolution is determined by the coupling of nodes through edges. For ease of notation, we will use the terms `station' and `node' interchangeably even though some nodes are actually junctions and not stations. We denote the delay of a station $i$ at time $t$ by $D_i(t)$. This variable is defined as the sum of the delays of all trains that are moving \textit{towards} station $i$ at time $t$:

\begin{equation}
\label{eq:Di}
D_i(t) = \sum_{T \in \mathcal{T}(i,t)} d_T(t), 
\end{equation}

where $\mathcal{T}(i,t)$ is the set of trains moving to station $i$ (i.e. the very next station they cross will be $i$, whether they stop there or not) at time $t$ and $d_T(t)$ denotes the delay carried by train $T$ at time $t$. We consider two ways in which the value of $D_i$ can change over time:

\begin{enumerate}
    \item A train, which was previously moving towards another station $j$, reaches $j$ and is now moving towards $i$. Therefore, its delay is now added to $D_i$.
    \item A train, which was moving towards $i$, reaches $i$ and either moves further towards another station or ends its trajectory. Therefore, its delay is removed from $D_i$.
\end{enumerate}

\begin{figure}[!h]
    \centering
    \includegraphics[width=.95\linewidth]{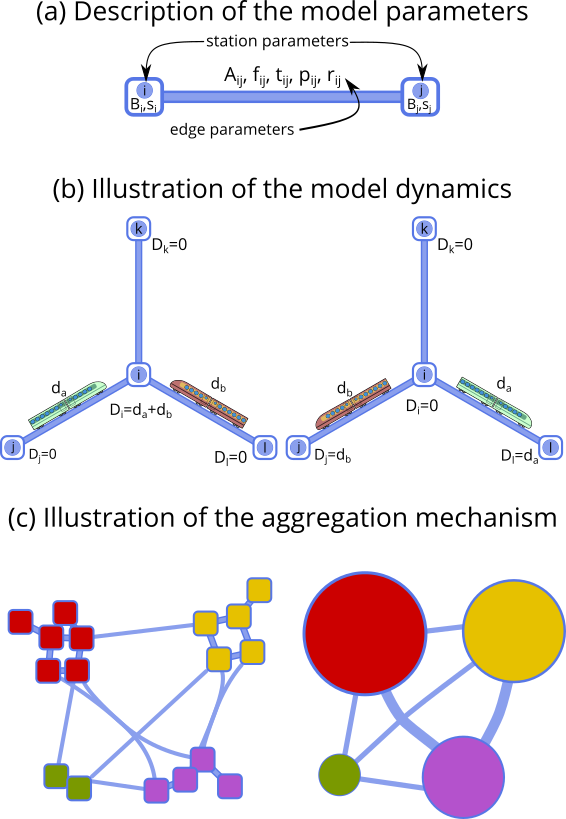}
    \caption{Model visualization: (a) station and line (edge in the network) dependent parameters, (b) illustration of the two mechanisms behind the delay dynamics dynamics, i.e. the appearance of new trains with their delays and the departure of the already included ones, (c) an example of the network aggregation. }
    \label{fig:2_illustration}
\end{figure}

The delay of station $i$ at the next time step --- we refer to this as $D_i(t+\Delta t)$, with $\Delta t$ being the time step size --- is dependent on the delays in various locations at the previous time step, not only $D_i(t)$. Thus, we write the relation between the delays between two consecutive time steps using a delay vector $\vec{D}= (D_1, D_2, \ldots, D_N)^T$, where $N$ is the total number of nodes:

\begin{equation}
\label{eq:full-eq}
D_i(t+\Delta t) - D_i(t) = \underbrace{F_{1,i}(\vec D)}_{\substack{\text{New incoming} \\ \text{trains towards $i$}}} - \underbrace{F_{2,i}(\vec D)}_{\substack{\text{Arrival of trains} \\ \text{at station $i$}}}
\end{equation}

with $F_{1,i}$ describing how the delay at station $i$ changes over a time step $\Delta t$ by means of the first term above (the addition of delay), and $F_{2,i}$ likewise by the second term above (the removal of delay). In the next section we express both these functions $F_{1,i}$ and $F_{2,i}$ in terms of several parameters and $\vec{D(t)}$. An illustration of the model and its terms is given in Fig.~\ref{fig:2_illustration}.

\subsection{Diffusion model equations}

The first term ($F_{1,i}$) sums the delays carried by all trains that start moving towards $i$ in the interval $[t, t+\Delta t]$:

\[
F_{1,i}(\vec D) = \sum_{\substack{\text{Trains $T$ that started moving to $i$} \\ \text{at $\tau\in[t,t+\Delta t]$}}} d_T(\tau).
\] 

This sum can be rewritten as a sum over the neighbors of $i$. By making a number of assumptions, like approximating the fraction of trains in each direction by the relative frequency (a full derivation can be found in Appendix~A), we can rewrite the delay of a train moving to a station $j$ as function of the delay of that station $D_j$ and express how many of the trains arriving at a station $j$ continue to $i$. This leads to:

\begin{equation}
\label{eqn:F1}
\begin{aligned}
F_{1,i}(\vec D) &=& \Delta t\sum_{j\in\mathcal{N}_\text{in}(i)}  p_{ji} B_j D_j(t).
\end{aligned}
\end{equation}

Here $\mathcal{N}_\text{in}(i)$ is the set of stations $j$ that have an edge to $i$. The parameter $p_{ji}$ is the probability that a train that reaches station $j$ will continue towards station $i$, and is computed as follows:

{\footnotesize
\begin{align}
p_{ji} &= P(\text{to }i|\text{from }j)\nonumber \\
&= P(\text{to }i|\text{(from }j\text{ \& do not end at }j\text{)})\cdot P(\text{do not end at }j)\nonumber \\
&= \frac{f_{ji}}{\displaystyle\sum_{\ell \in\mathcal{N}_\text{out}(j)} f_{j\ell}} \cdot\left(1- \substack{\text{Probability that train}\\\text{has end station at $j$}}\right)\nonumber \\
&= r_{ji} (1-s_j) \label{eq:pij},
\end{align}}

where $\mathcal{N}_\text{out}(j)$ is the set of stations to which there is an edge from $j$. The value of $p_{ji}$ is equal to a multiplication of two factors. The first (denoted by $r_{ji}$) is the probability that if a train reaches $j$ and it does not end its journey there, it will then continue towards $i$. Note that we consider this probability to be independent of where the train came from: we do not consider any memory in this process. The value is calculated as the frequency of  trains going from $j$ to $i$ divided by the frequency of all outgoing trains from $j$. 
The second factor in Eqn.~\ref{eq:pij} (denoted by $1-s_j$) is the probability that the train does not end at $j$ --- $s_j$ itself is the probability that a train that arrived at $j$ ends its journey there, for example because it is the terminus. The variable $B_i$ in Eqn.~\ref{eqn:F1} is a station-dependent parameter, defined as

\begin{equation}
    \label{eq:Bi}
    B_i = \frac{\sum_\text{edges $e$ to $i$} f_e}{\sum_\text{edges $e$ to $i$} f_e t_e}=\frac{\sum_{\ell\in\mathcal{N}_\text{in}(i)} f_{\ell i}}{\sum_{\ell\in\mathcal{N}_\text{in}(i)} f_{\ell i} t_{\ell i}},
\end{equation}

where $f_e$ denotes the frequency of trains on edge $e$, and $t_e$ corresponds to the time a train takes to cross edge $e$. The parameter $B_i$ has units of time$^{-1}$ and can therefore be interpreted as a rate. The inverse of $B_i$ is the average time of edges towards $i$, weighted by their frequency. A high value of $B_i$ corresponds to a station with incoming short edges with high frequency. Intuitively, $B_i$ can be thought of as a station's train turnover rate.

The second term of Eqn.~\ref{eq:full-eq} ($F_{2,i}$) counts the delays of trains that reach station $i$ and therefore remove their delays from $D_i$. We express $F_{2,i}$ as follows (for details see Appendix~A):

\begin{align}
F_{2, i}(\vec D) &= \sum_{\substack{\text{Trains $T$ that reached $i$} \\ \text{at $\tau\in[t,t+\Delta t]$}}} d_T(\tau)\nonumber \\
&= \Delta t B_i D_i(t). \label{eqn:F2}
\end{align}

The term only depends on the delay $D_i(t)$ at station $i$ at the previous time step, and the previously mentioned parameter $B_i$. The delay loss at a station can be interpreted as an exponential process with rate $B_i$.

The contributions $F_1$ and $F_2$ are expressed in terms of the delay state vector $\vec{D}$ and in terms of various railway parameters (summarized in Tab.~\ref{tab:paramdesc}). Filling in these two terms into Eqn.~\ref{eq:full-eq} gives the full expression for the evolution of the delay $D$ at any station $i$:

{\footnotesize
\begin{align*}
 D_i(t+\Delta t) - D_i(t) = \Delta t\left[\sum_{j\in\mathcal{N}_\text{in}(i)}  p_{ji}B_j D_j(t) - D_i(t) B_i\right].
\end{align*}}

We can simplify the sum over the neighbours of $i$ by using the railway network's adjacency matrix $A$, which has entries entries $A_{ji} = 1$ if there is an edge from station $j$ to station $i$ and entries zero elsewhere:

\begin{equation*}
\label{eq:fineq}
\frac{D_i(t+\Delta t) - D_i(t)}{\Delta t} = \sum_j A_{ji} B_j D_j(t) p_{ji} - D_i(t) B_i.
\end{equation*}

Here, the sum goes over all nodes $j$. This equation can be written in matrix form using $\vec{D}$ as a column vector. Moreover, we can take the limit $\Delta t\rightarrow 0$. This leads to the expression

\begin{equation}
\label{eq:difeq}
\frac{d \vec{D}(t)}{d t} =  \mathbf{G} \cdot \vec{D}(t).
\end{equation}

The above equation contains the core model matrix $\mathbf{G}$, an $N\times N$ matrix defined as follows ($\delta_{ij}$ is the Kronecker delta):

\begin{equation}
\label{eq:GH}
G_{ij} = A_{ji} p_{ji} B_j - \delta_{ij} B_j.
\end{equation}

All of the dynamics of the model are encapsulated in the matrix $\mathbf{G}$.


\begin{table*}[!h]
\centering
\begin{tabular}{ll}
\hline
Variable                    & Description                                     \\ \hline
$D_i(t)$                    & Delay at station $i$ and time $t$                 \\
$A_{ji}$                    & Adjacency matrix of the railway network         \\
$B_i$                       & Train turnover rate of station $i$                                          \\
$d_T(t)$                    & Delay carried by train $T$ at time $t$          \\
$f_{ij}$                    & Train frequency from stations $i$ to $j$        \\
$\bar{t}_{ij}$              & Average travel time from station $i$ to $j$     \\
$p_{ji}$                    & Fraction of trains to $j$ that continues to $i$ \\
$r_{ji}$                    & Fraction of trains to $j$ that continue to $i$ if they do not end at $j$ \\
$s_j$                       & Fraction of trains that end at station $j$         \\
$\mathcal{T}(i,t)$          & Set of trains moving to station $i$ at time $t$ \\
$\mathcal{N}_\text{out}(j)$ & Set of stations to which there is an edge from $j$             \\
$\mathcal{N}_\text{in}(j)$  & Set of stations from which there is an edge towards $j$              \\
$N$                         & Amount of stations \\
$\delta_{ij}$               & Kronecker delta ($\delta_{ij}=1$ if $i=j$, and $0$ otherwise) \\
\hline
\end{tabular}
\caption{Overview of the model variables and parameters.}
\label{tab:paramdesc}
\end{table*}

\subsection{Model aggregation to clusters of stations}
\label{sec:model_agg}

In this paper, we aim to describe how well our model describes real delay propagation patterns. One variable in this analysis is the level of spatial aggregation at which we simulate the model. In the previous section we explained the model where each node of the network consists of a single station or junction. However, the same principles can be applied to a network where nodes correspond to a group of such stations. The method we use to group stations into clusters is explained in Sec.~\ref{sec:agg}. Here, we discuss how the model parameters for the full-resolution model based on individual stations can be translated into a lower resolution version. The discussed aggregation process is very similar to  network of networks idea known in the networks literature \cite{Gao2011, Kivela2014, Siudem2019}.

Above, each delay variable $D_i$ corresponds to one node of the network. This is achieved by transforming delays on trains to delays on stations via Eqn.~\ref{eq:Di}. This is already a form of coarse-graining the delay dynamics. Now, we assume that the original railway network of $N$ stations is divided into $K$ clusters (or groups of stations). We indicate stations with lowercase letters ($i$ and $j$) and clusters with uppercase letters ($I$ and $J$). The clusters naturally form a network: an edge between clusters $I$ and $J$ exists if there is at least one station $i$ in $I$ and one $j$ in $J$ such that there is an edge between $i$ and $j$ in the original network. We define the function $\mathcal{C}$ from stations to clusters such that $\mathcal{C}(i)$ is the cluster to which station $i$ belongs. Let $D_I(t)$ denote the total delay of all trains moving to any station in cluster $I$ at time $t$, either from inside the cluster, or coming from other clusters. An equation for the evolution of this delay can be derived in the same way as we did above for stations. The delay $D_I$ can change when trains start towards any station in this cluster, or when trains arrive at a station in this cluster. The main difference with the non-clustered case (above) is the fact that in the clustered case, self-loops in the network appear. This is because trains moving to a station in a cluster --- and thus adding to the cluster's delay --- can reach that station, and then continue to another station in the same cluster, again adding to the cluster's delay. 

While the equations in the clustered case are the same as in the non-clustered case, the parameters such as frequencies and travel times are now defined on edges between clusters. We explain a method to express these cluster parameters in terms of their non-clustered counterparts (i.e., the ones in Tab.~\ref{tab:paramdesc}). We start with the total frequency $f_{IJ}$ and weighted averaged travel time $t_{IJ}$ of trains between two clusters $I$ and $J$. We define them as

\begin{align}
    f_{IJ} &= \sum_{i\in I}\sum_{j\in J} f_{ij} \\
    t_{IJ} &= \frac{\sum_{i\in I}\sum_{j\in J} t_{ij}f_{ij} }{\sum_{i\in I}\sum_{j\in J} f_{ij}}.
\end{align}

These definitions are intuitive: the total frequency of trains between two clusters is the sum of the frequencies on edges going from a station in the first to a station in the second cluster. The travel time is the weighted average of the travel times of the edges going from the first to the second, weighted by their frequency. 

Next, we need to define the stopping probability $s_I$ for a cluster $I$. In order to do this, we define the station parameter $q_i$ as the probability that a train which arrives in the cluster $\mathcal{C}(i)$, arrives at station $i$. In a way, it indicates how important station $i$ is in its cluster, measured by the total frequency of all incoming trains to that station. The quantity is approximated as follows:

\begin{equation}
    q_i = \frac{\sum_{j \in\mathcal{N}(i)} f_{ji} }{\sum_{j \in \mathcal{C}(i)}\sum_{\ell\in\mathcal{N}(j)} f_{\ell j}}.
\end{equation}

Note that the values $q_i$ are weights of stations whose sum is one. Each station is weighted by the frequency of incoming trains. Next we use the quantities $q_i$ to estimate the stopping probability $s_I$ for cluster $I$:

\begin{equation}
    s_I = \sum_{i\in I} s_i q_i.
\end{equation}

One can interpret the stopping probability formula using the following formula:

{\footnotesize
\begin{equation*}
    s_I = \sum_{i \in I} P(\text{stops in }i|\text{arrived in }i) P(\text{arrives in }i|\text{arrives in }I)
\end{equation*}}

Using this approach, we can set up a model for any clustering of the original network using only the parameters of the full network. We can thus compute the matrix $\mathbf{G}$ (Eqn.~\ref{eq:difeq} and \ref{eq:GH}) for each clustered case. We denote such matrices of the clustered model by $\mathbf{G}_c$.

An additional possibility, which we do not discuss further, is to define a clustered model directly, without relying on the parameters of the full network. In this case, the frequencies, average travel times and stopping probabilities need to be directly measured from data. 
\begin{figure*}[!ht]
	\centering
	\includegraphics[width=2\linewidth]{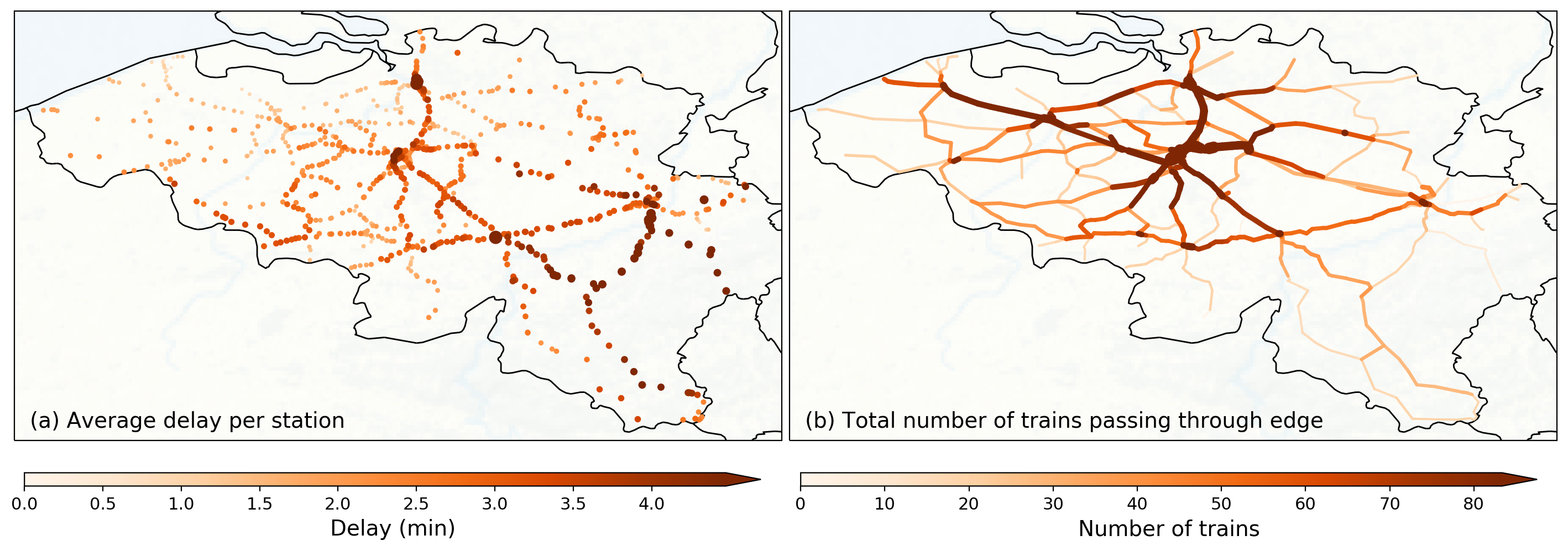}
	\caption{\textbf{Panel (a)}: Average delay per train in November 2019, shown at every node. \textbf{Panel (b)}: Average number of trains passed through the edge on April 11th, 2019 (taken as an example day). Only passenger trains are used when calculating these numbers.}
	\label{fig:2_averages}
\end{figure*}

\subsection{Model considerations}

There are a number of important assumptions we used in our model (see also Appendix~A). Because we aggregate delays from trains onto stations, we lose a lot of details, such as origin-destination information of trains. In the derivation of the model, a delay `arriving' at a station is subsequently spread out and propagated to all neighbors of that station, based on a fixed weighting of the outgoing edges. However, in real railway systems, there is a high correlation between where the delay comes from and where it goes to, and memory effects can be important. Our model is expected to work better on lower spatial resolution, on scales where a lot of trains and train routes contribute to the dynamics of a single node, such that trains picking a random direction constitute a decent approximation to the real dynamics, which on the detailed level is inherently schedule-based and not random. Furthermore, the delays in our model are treated as variables smoothly varying in time and space. In reality, delays which are localized in space are of a discrete nature: a single train can be delayed, and when the train has `passed' a station, the delay suddenly disappears from this station. This means that the time series of $D_i(t)$ in reality has a lot of jumps, namely every time a train reaches this station or starts towards it. In the model, $D_i(t)$ is smoothly varying. Another important consideration is that the model only propagates delay and removes delay from the system --- it does not add any new delays. Moreover, the only mechanism by which delays are removed from the station is when a train ends its trajectory, which is encoded in the parameters $s_i$. An underlying assumption is thus that each train keeps its initial delay until it has reached its final stop. In practice, of course, delays are constantly generated, often due to small noise-like incidents or other (delayed) trains blocking platforms or tracks, and in more exceptional cases due to new disruptions. Moreover, trains can lose some delay by traveling faster or because of scheduled buffer times at stations, which is not included in our model. For these reasons, we will only compare the results of our model to data of days with a large disruption: by focusing on a time point with a large amount of delay and analyzing its dissipation through the network, we minimize the effects of smaller stochastic delays, which are expected to contribute less to the dynamics in these situations. A final limitation we would like to mention is that in our model, the finite travel time of trains and their location on an edge is lost: in our assumptions, a train's delay counts fully towards to the next station's delay, wherever the train is on an edge towards that station. For small time steps, this means the train's delay also counts immediately to the propagated delay further on in the network, even if in reality the train would still need more time to cross the edge.

Next to the limitations mentioned here, our model also has clear benefits: a compact description (the matrix $\mathbf{G}$), the fact that it is linear and thus amenable to analytical study and the straightforward generalization to lower spatial resolution. We discuss advantages of the model throughout, and at the end of this paper.
 
Some of the limitations mentioned above directly stem from our choice for a network-based, diffusion-like model. It is one of the aims of this paper to investigate whether our model, and its built-in potential for spatial aggregation, can reproduce the dynamics of delay propagation observed in a real railway system. 

\section{Data and Methods}
\label{sec:3_Application}

We apply the model to the Belgian railway system as an example. We chose the Belgian railway system for multiple reasons. Being a West-European country, Belgium has a rather dense and strongly utilised railway system with over 100 km of lines per 1000 m$^2$, being one of the world's densest national railway systems \cite{IUR}. In contrast to, for example, the American or Chinese railways (both have about 10-25 km of railways per 1000 m$^2$). Additionally, freight and high-speed trains make up only a small fraction of the total railway transport in this country. These aspects require more complex scheduling in the Belgian case, and it implies a more interesting delay evolution to use as an example. Another reason for analysing the Belgian railways is the availability of data, which is discussed below. A discussion on the international relevance of the results is given in the conclusions.

\subsection{Data and Pre-processing}

We use the open data provided by Infrabel, the service company of the Belgian railway network \cite{Infrabel}. The data contains geographical information on railway stations and the physical railway lines, recorded tracks of passenger trains with details on scheduled and realised departure and arrival times of their activities on each station or junction, as well as associated delays. The time stamps and delay data are in seconds. We use data from all Belgian passenger railway activities between January 2019 until May 2020. The data covers an average number of 3600 daily unique trains on business days, and 2200 on weekends or holidays.

The first step is to reconstruct the graph of the Belgian railway network. First we add all stations as nodes in our graph. We get the edges by mapping the geographical locations of railway stations onto geographical shapes of railway lines and every two stations are connected together iff there is a line connecting them without intermediate stations. The geometry of railway lines is more intricate than simple edges between stations, since there exist places of splits and merges of multiple lines. We implement these by adding so-called “junction” nodes along the lines. 

The dataset contains all railway stations, which except of passenger train stations include merchandise platforms, technical depots, carwashes, etc. Passenger trains tend to skip those intermediate platforms and the passage information is not recorded. In order to bypass this limitation, in case when there is no edge between two consecutive stations in the track record, we assume that the train follows the shortest path between them. Delay accumulation or reduction is then evenly spread across the intermediate stations along that path.

There are two kinds of passenger trains available in the data that can be characterised by the proportion of skipped stations along the track: 1) local trains, which usually circulate at shorter distances and stop at every station along the path and 2) intercity trains, which circulate at larger distances and skip some intermediate stations. We exclude from the analysis the intercity trains that skip a significant portion of stations along the track (usually these are international trains) and extra trains that run ad hoc on a specific day. The amount of disregarded trains is less than 3-5\% of the total data. We further use the notion of a railway graph and a railway network interchangeably.

The reconstructed network and two important delay statistics are shown in Figure~\ref{fig:2_averages}. The graph contains 822 stations and 972 edges. Because the network has mostly a line-like structure, 78\% of all stations have degree 2 and the average degree is 2.19. In panel Fig.~\ref{fig:2_averages}a, we show the average delay of trains travelling towards stations in November 2019. A general trend from small average delays in the north-west to larger average delays in the south-east is visible, with the cities of Antwerp (north) and Brussels (centre, the capital) also having rather high average delays. Panel Fig.~\ref{fig:2_averages}b colours the edges of the network with the average amount of trains per day that crosses them. Several lines between the large cities of Bruges, Ghent, Brussels and Antwerp stand out.

We use the recorded tracks to estimate the model parameters. In particular, we calculate the edge parameters $f_{ij}$ and $t_{ij}$ and the station parameters $s_i$ (see Table \ref{tab:paramdesc}) for each month separately. Within a month we aggregate all frequency and temporal counts for each day of the week. Moreover, for each day we keep separate counts for six 4-hour periods of the day. For each station $j$ this leads to the estimation of parameters $s_j$ as the average fraction of arriving trains that end their trajectory at station $j$, and $f_{ij}$ and $t_{ij}$, the average frequency and average  passage time of trains going from station $i$ to $j$. For simulations of disrupted situations, we use the parameters obtained for the month, day of the week and period of the day corresponding to the timing of the peak delay on the disrupted days. If not mentioned otherwise, we use $\Delta t = 30$ seconds in all results in this paper. Simulations of the model were coded in Python. For the clustering discussed in 3.4, we used the KMeans function of scikit-learn. The data and code is publicly available and we refer the reader for this to the appropriate section at the very end of the paper.

\subsection{Disrupted situations}
\label{sec:disr_sit}

As discussed in the introduction, we expect diffusion-like models to be of most interest to study large-scale delay propagation: e.g., general directions of delay evolutions --- as individual delays will be predicted erroneously due to lack of trajectory information of delays by abstracting away from individual trains. Therefore, we focus our model analysis on days in which such large-scale delay propagation can be assumed important, namely where the delays were severe. In contrast, when delays are small, they dissipate quickly and will not spread much --- making identifying large-scale spread of delay of less interest. Another reason why we focus on days with severe delays is that understanding such days is of great importance to railway companies to be able to handle such situations well. We refer to days with severe delays as `disrupted days'. A list of disrupted days is obtained by looking at the peak in the total delays (i.e., delay summed over all nodes at any given moment in time) of every day in the dataset, and taking the 50 days with the highest peaks. The exact dates in this list are given in Appendix~B. Throughout the rest of the paper, we initialize our simulations at the peak in total delay of these disrupted days, i.e. we determine the delay on each station at the time of peak delay and use this as initial vector of delays. The model captures the spread and dissipation of existing delays, so these simulations will capture the propagation of the delays present at the peak --- making such an initial point most interesting. Also, we reason that after the moment of highest total delay, the relative importance of newly generated (i.e., non-captured) delays is small as compared to existing delays.

\begin{figure*}[!h]
    \centering
    \includegraphics[width=2\linewidth]{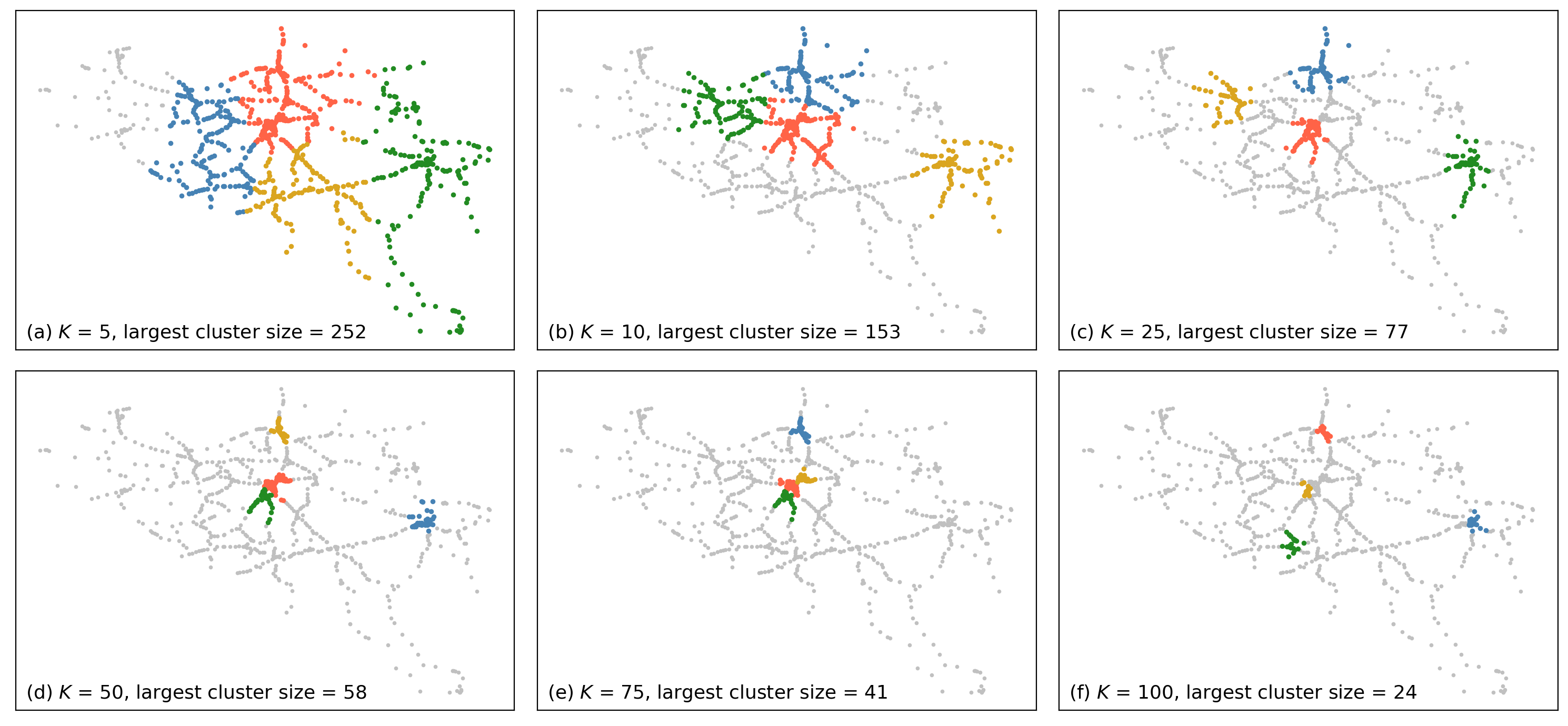}
    \caption{Clustering results for (a) $K=5$, (b) $K=10$, (c) $K=25$, (d) $K=50$, (e) $K=75$ and (f) $K=100$. For visibility purposes, only the fourth largest clusters are coloured (in the order of red, blue, green and yellow). Stations not belonging to any cluster are coloured grey. Cluster size is measured by amount of associated stations. Largest cluster sizes are denoted in panel labels.}
    \label{fig:3_clusters}
\end{figure*}

\subsection{Quantifying model performance}
\label{sec:modelperf}

When assessing the model's performance to reproduce reality, we focus on whether the model reproduces the correct direction of delay evolution, rather than simulating exact values well. There are a number of reasons for this. First, when aiming to understand large-scale propagation of severe delays --- which is the aim of this model --- accurately tracking the positioning of delays (rather than the exact values) through space is already very important information to practitioners. Analysing such directional trends of delays provides us with information on how the system works, absolute values of delays are not always necessary for that. Second, in severely delayed circumstances, numeric performance comparison can quickly become biased by several high spikes in delays: particular trains being up to one hour delayed, compared to an average delay of a couple of minutes in the rest of the network. And third, our model was not designed to capture small, stochastic variations in the delays. However, such delays are always present in the data, which means that one will never get a good quantitative fit, even if the model would be a perfect representation of the propagation of existing delays. For these reasons, we use Spearman's correlation coefficient $\rho$ to measure the model's performance. This metric is based on the rank of the variables, i.e., it assess monotonic relationships rather than linear relationships (which is the case, for example, for Pearson's correlation coefficient). We denote the observed delays at all stations at time $t$ by $\vec{D}_{obs}(t)$: a vector with delay entries per station. Likewise, we define a simulated delay vector $\vec{D}_{sim}(t)$. We denote the vector containing the ranks of the stations based on their delays by $r(\vec{D}(t))$, using either observed or simulated delays. Then, Spearman's correlation coefficient at time $t$ is given by:

\begin{eqnarray*}
\label{eqn:spearman}
\rho(t) &=& \text{Pearson}\bigg( r(\vec{D}_{obs}(t)), r(\vec{D}_{sim}(t)) \bigg) \\
&=& \frac{\text{cov}[r(\vec{D}_{obs}(t)), r(\vec{D}_{sim}(t))]}{\sigma_{r(\vec{D}_{obs}(t))}\sigma_{r(\vec{D}_{sim}(t))}}
\end{eqnarray*}

In the next section, we use this metric to compare the model performance on different levels of spatial aggregation. In the clustered case, the vector $\vec{D}_{sim}(t)$ will not have $N$ elements (the number of stations), but $K<N$, the number of clusters. Each component of the vector is the total delay in one cluster. We want to compare this with the observed data on $N$ stations. This observed delay vector therefore also has to be aggregated on the $K$ clusters, by summing the delays of the stations belonging to the same cluster. To be able to compare the Spearman correlations across simulations with different $K$, the observed and simulated delay vectors of dimension $K$ are de-aggregated towards dimension $N$ (i.e., equally distributed across each cluster's stations), such that we always compute Spearman's correlation on vectors of length $N$, even if the model was simulated on the network of clusters.

\subsection{Clustering}
\label{sec:agg}

When referring to `level of aggregation', we mean the spatial resolution of the model. Full resolution would mean using all stations as entities in the model (i.e., no clustering), and lower resolutions involve clusters or groups of stations as entities in the model. Section~\ref{sec:model_agg} describes how we translate node and edge parameters towards a lower resolution. Here we discuss the means of clustering itself: the process of grouping stations in an appropriate manner. Many of such clustering methods exist, and we have chosen to use K-means \cite{Steinhaus1956, MacQueen1967} on the spatial coordinates of the nodes in the Belgian railway network (longitude, latitude). We do this for the following reasons. First of all, we aim to create groups of stations that are adjacent to each other. Although the way we use K-means does not explicitly incorporate network topology, it does make sure that the groups of stations are convex (i.e., there is no station from cluster $A$ in the middle of cluster $B$), since the railway network is an inherently spatial network. This geographic basis for the groups also makes them easier to interpret. Another important reason for using K-means is that we can choose $K$ --- the desired number of clusters --- which we can vary to get different levels of aggregation to assess the model performance with.

We vary $K$ between a minimum amount $K_{min}$ and maximum amount $K_{max}$ of clusters, which in this case we set to be 3 and 100, respectively. Note that values of $K_{min}$ lower van 3 are excluded because of the resulting coarseness of the resulting model, and values of $K_{max}$ higher than 100 are excluded because they result in many single-station clusters. The K-means algorithm starts with an initial set of $K$ points (`centroids') and assigns all stations to the closest centroid. Each centroid now corresponds to a cluster of stations. Next, the centroid coordinates are redefined as the average of all the stations in its cluster. This process is then iterated (reassigning stations to closet centroid, updating centroid coordinates) until it converges to a point where the centroids do not change anymore.

The resulting clusters for several values $K$ are shown in Fig.~\ref{fig:3_clusters}. In each plot, the four largest clusters in the network are shown in colours. Observe the small size of clusters in the $K=100$ case, motivating the $K_{max}=100$ threshold. We can also see that the largest clusters (in terms of number of stations) for high values of $K$ are situated around the major cities of Brussels, Antwerp and Li\`ege. This can be explained by the fact that these cities contain numerous smaller railway stations that are geographically close together, while in more rural areas like the south and west, the station density is much smaller. Urban areas are thus expected to contain larger clusters for relatively large values of $K$.

\begin{figure*}[!h]
    \centering
    \includegraphics[width=2\linewidth]{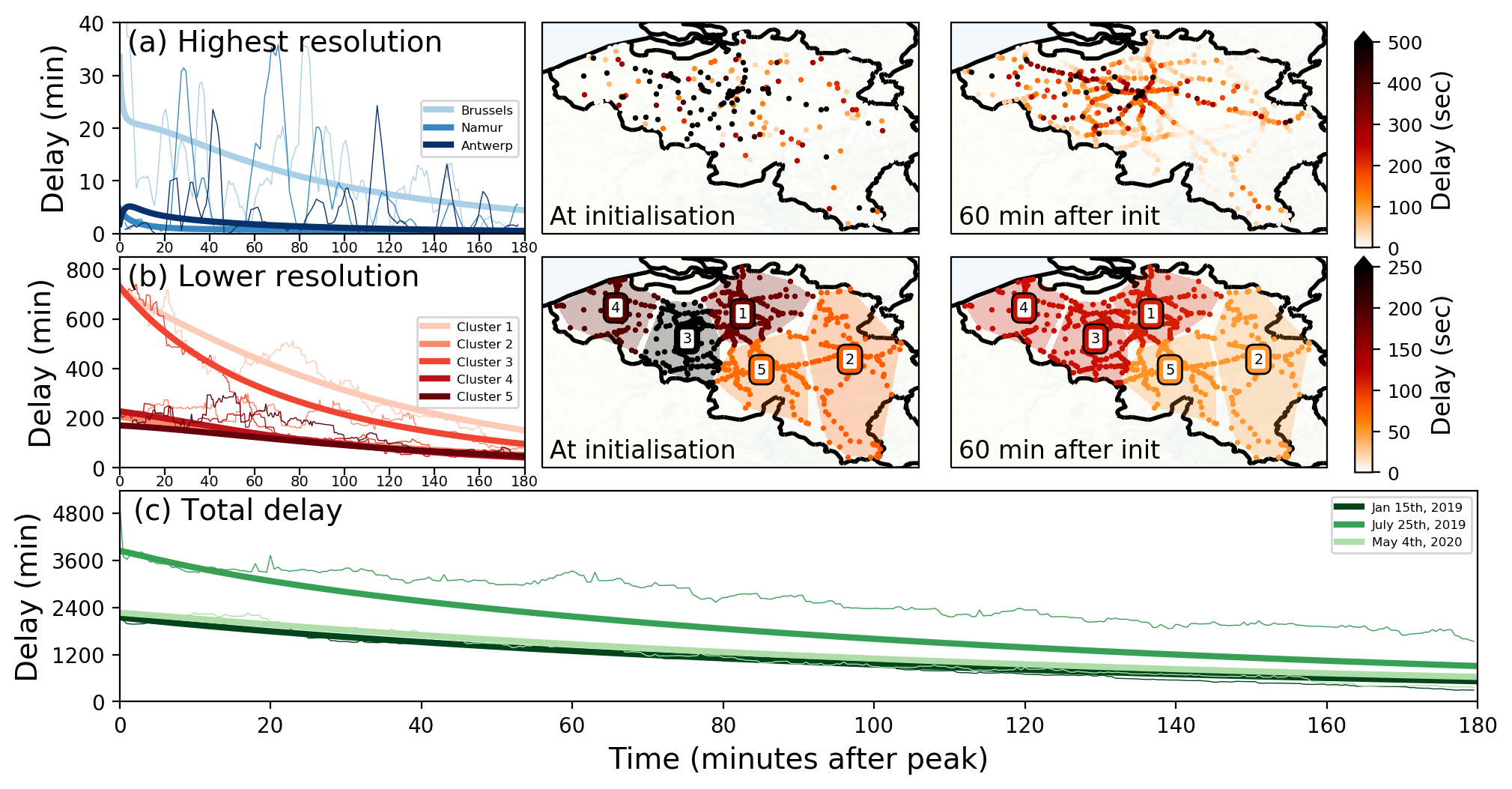}
    \caption{Example simulations at various resolutions. \textbf{Panel (a)}: highest (non-clustered) resolution simulation of Jan 15th, 2019, initialised at the peak delay (18:11). The delay evolution over time of three example stations is displayed in blue lines (Brussels, Namur and Antwerp). The spatial situation of simulated delays at initialisation (middle) and 60 min after initialisation (right) are also shown. \textbf{Panel (b)}: Simulation of the same day, but at an spatially aggregated level of five clusters. Red lines in the left panel show the temporal evolution of delays for each cluster. Again, the middle and right panels indicate spatial delay distribution at initialisation and 60 min after initialisation. For clarity, the clusters are shaded in the background. \textbf{Panel (c)}: Total delay evolution of three example bad days: Jan 15th, 2019, July 25th, 2019 and May 4th 2020, all initialised at their peak delay moments. All maps show delays in seconds, with a cut-off at 500 and 250 seconds respectively, as higher delays were rare on these instances.}
    \label{fig:4_examples}
\end{figure*}

\section{Results}
\label{sec:4_Results}

In this section, we show the dynamics of the model and compare it with the data for a number of disrupted days as example. Then, we apply the model to toy examples to illustrate in which circumstances this model works well and in which it does not. We end by discussing the overall performance on all 50 disrupted days.

\subsection{Example simulation}

We start by looking at a few example disrupted days. We start with Jan 15th, 2019, which had a peak delay at 18:11. Initialising the non-clustered (`highest resolution') version of the model at this moment, we simulated the delays up to three hours after the peak. The delay evolutions at three major stations (Brussels, Namur and Antwerp) are displayed in Fig.~\ref{fig:4_examples}a. It is clearly visible that the simulated delay time series is much smoother than the real time series, which has strong jumps as a consequence of the discrete nature of trains: either delayed trains are going to those stations (i.e. delay $>0$), or not (i.e., delay $=0$). This is also visible in the maps in the upper row of this figure: at initialisation time, the delays are distributed very discretely across the network (center-top panel). The model diffuses the delay across the network after 60 min (right-top panel). In this figure, we can clearly see one assumption on which the diffusion-like model is based: it assumes that delay is spread by a very large amount of trains, and that it travels to all other adjacent stations instantly (albeit weighted into small fractions). Of course, in reality this assumption does not hold.

In panel (b) of Fig.~\ref{fig:4_examples} we take the exact same day, but instead of modelling at full resolution, we cluster the network into five clusters and redo the analysis. We observe that, by aggregating over the many trains present in each cluster, the jumps in delays visible in panel (a) become less pronounced: the real delay evolution curves per cluster in panel (b) are more smooth. The general trends of the real delay curves in each respective cluster resembles the simulation quite well, even though some undulations are visible. For example, increases in delays in cluster 1 around 70 minutes and in cluster 5 around 40 minutes after initialisation are visible. This is the result of newly generated delays. Snapshots of the spatial distributions of the delay are shown in the maps on the right. They show how delay dissipates and is transported across the five clusters. A quick comparison by eye between the highest resolution maps (top panels) and these lower resolution maps indicates the resemblance.

Panel (c) of Fig.~\ref{fig:4_examples} shows the evolution of the total delay in the system (both simulated and real) in three example disrupted days: Jan 15th, 2019, July 25th, 2019 and May 4th 2020. One can see that the total delay on Jan 15th and May 4th are simulated quite well over the whole three hours, but the real total delay on July 25th quickly overshoots the simulated curve --- pointing towards the effect of newly generated delays.

\begin{figure*}[!h]
    \centering
    \includegraphics[width=2\linewidth]{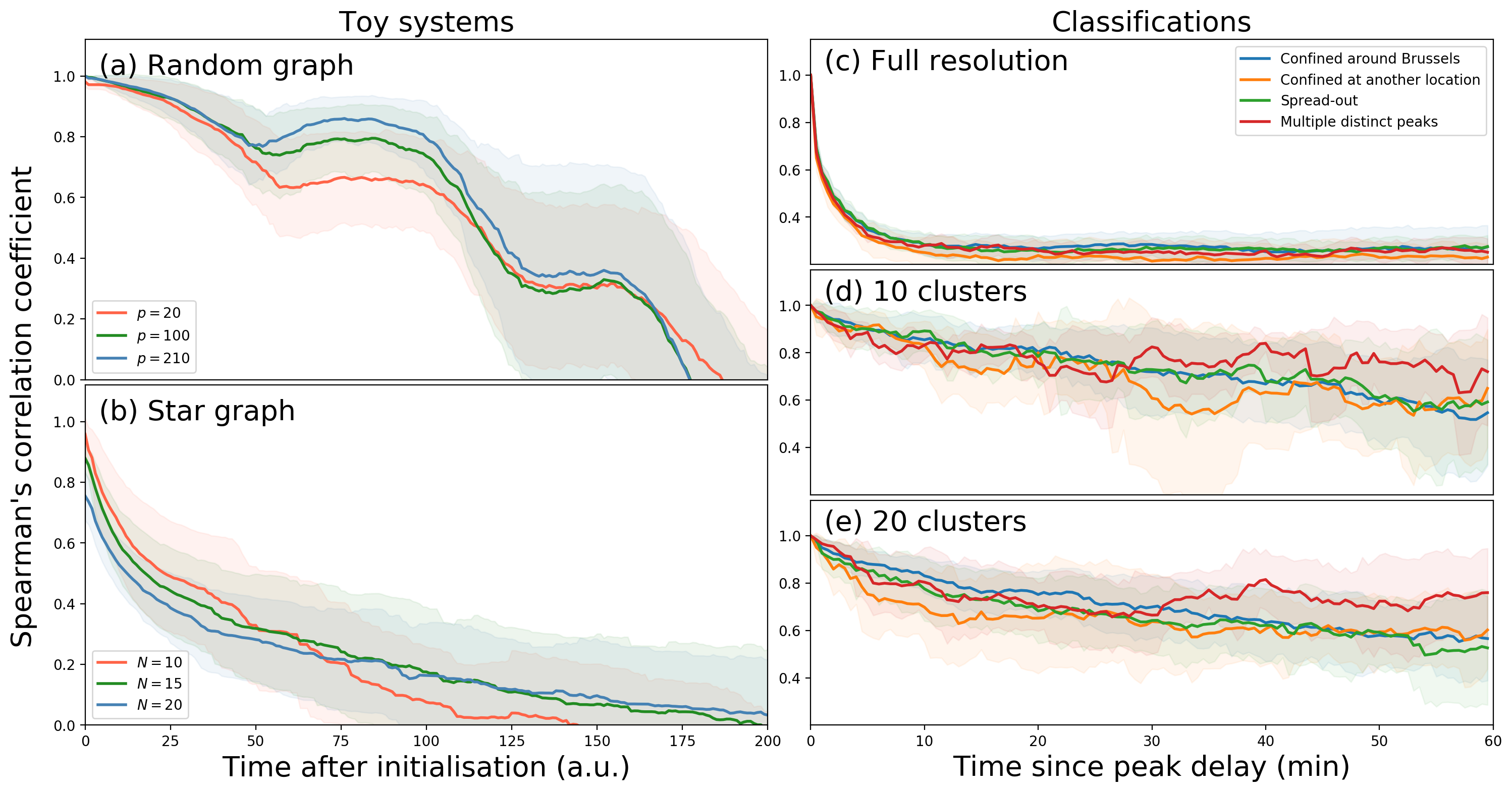}
    \caption{Model performance in toy systems and across classifications of disrupted days. \textbf{Panel (a)}: Model performance of the random toy system for three different values of the amount of lines $p$ (see Appendix~D for details). \textbf{Panel (b)}: Model performance of the star graph toy system for different values of the amount of nodes $N$ (keeping $p$ fixed, see Appendix~D for more details). \textbf{Panel (c)-(e)}: Model performance along time since peak delay of the disrupted days, averaged within each classification (see Appendix~B for details), for the model at (a) full resolution, (b) using $K=10$ and (c) using $K=20$ clusters. Averages are shown in lines, shaded areas indicate the range of one standard deviation from the average.}
    \label{fig:5_scenarios}
\end{figure*}

\subsection{Toy model}

In this subsection we introduce two toy systems that allow us to study more fundamental properties of the model from section 2. They are explained in more detail in Appendix~D. The toy systems represent implementations of the model for networks with very simple topologies: random networks and star networks. Numerous other toy systems can be thought of, but we specifically compared these because they can test the model performance under different levels of the density of lines and connectivity of nodes. As for the real data we measure the model performance using the Spearman's rank correlation coefficient.

Fig.~\ref{fig:5_scenarios}a shows the performance of the model on a random graph-topology (with 15 nodes and 20 edges). We vary the amount of lines $p$ from 20, to 100, up to 210 (which is the maximum number of unique pairs in a 15-node connected graph). It is clear that the model performance decreases over time. At first, the different values of $p$ do not matter: the model performance decreases slightly due to the fact that in the model, delays are instantly spreading to various directions further in the system, while in these systems (and in reality), delays need to arrive at next stations first (carried by trains) before moving onto next stations. This leads to a discrepancy. As soon as the first trains arrive at next stations (around $t=50$), their delays contribute to delays on new edges where the model already predicted a small part of it to be. For a small amount of lines (i.e., low $p$), the specific direction the train is going is very important. For large amounts of lines (i.e., high $p$), all combined `chosen' directions of the trains approximate the frequency and other attributes put in the model. In other words, the model approximates reality better for densely used lines. And this seems to be visible: high $p$ (blue line) starts deviating positively from the red line after $t=50$. At much later points, the initial delays start arriving at their ending stations, which collapses the correlation down to much lower values.

For the star graphs (see panel (b) of Fig.~\ref{fig:5_scenarios}), where we fix the number of lines $p=50$  we take a look on the dependence on the number of nodes. The number of nodes do not seem to matter much, but it is clear that the star graph indicates much smaller correlations than the random graph. Although this is merely an example system, we intuitively expect that as soon as trains start driving towards the center, other delays (as a consequence of the diffusion-like nature of the model) are simulated to be at each of the connected nodes, quickly limiting the correlation. 

The above toy systems reflect that our model works better for denser networks with the higher number of train lines.

\subsection{Classification of disrupted days}

We now investigate whether the initial geographical delay distribution has an effect on the accuracy of the diffusion-like model. For this, we distinguish four categories across the 50 disrupted days, classified by eye based on the delay patterns on the peak delay moments. Appendix~A discusses this classification in more detail and also shows the delay maps. The first and largest group (25 days) contains the situations where almost all of the delays are localized near Brussels, the capital city of Belgium and important railway hub. In Belgium, train lines between east and west and north and south respectively all pass through Brussels, which makes it an important factor in the delay dynamics in the railway network. The second group (7 days) contains situations where the delay is also localized, but on a different location than Brussels. The third group (5 days) contains those situations with multiple locations with high delays. Finally, we consider the group of stations (13 days) where the delay is not localized but instead spread out over a large region. 

As before, we perform simulations with as initial condition the peak delay distribution. In Fig.~\ref{fig:5_scenarios}c-e we show the evolution of the Spearman correlation over time, averaged per group. We show this for different spatial resolutions. We find that there are no clear differences between the groups. The situations with delays localized near a city which is not Brussels (shown in orange) seem to perform a bit worse than the others, but we should be very cautious in interpreting this: the variation within a group is very large, as shown in the shaded areas.

The fact that there is no clear difference in model performance between groups could indicate that the spatial localization of the disruption is not a good determinant of the accuracy of our diffusion-like model. The obvious question then is: is there a better measure, or characteristic, which can distinguish between different disrupted situations and indicates whether a diffusion-like delay spread is warranted? We plan to explore this in future work. 

\subsection{Overall performance}

\begin{figure*}[!h]
    \centering
    \includegraphics[width=2\linewidth]{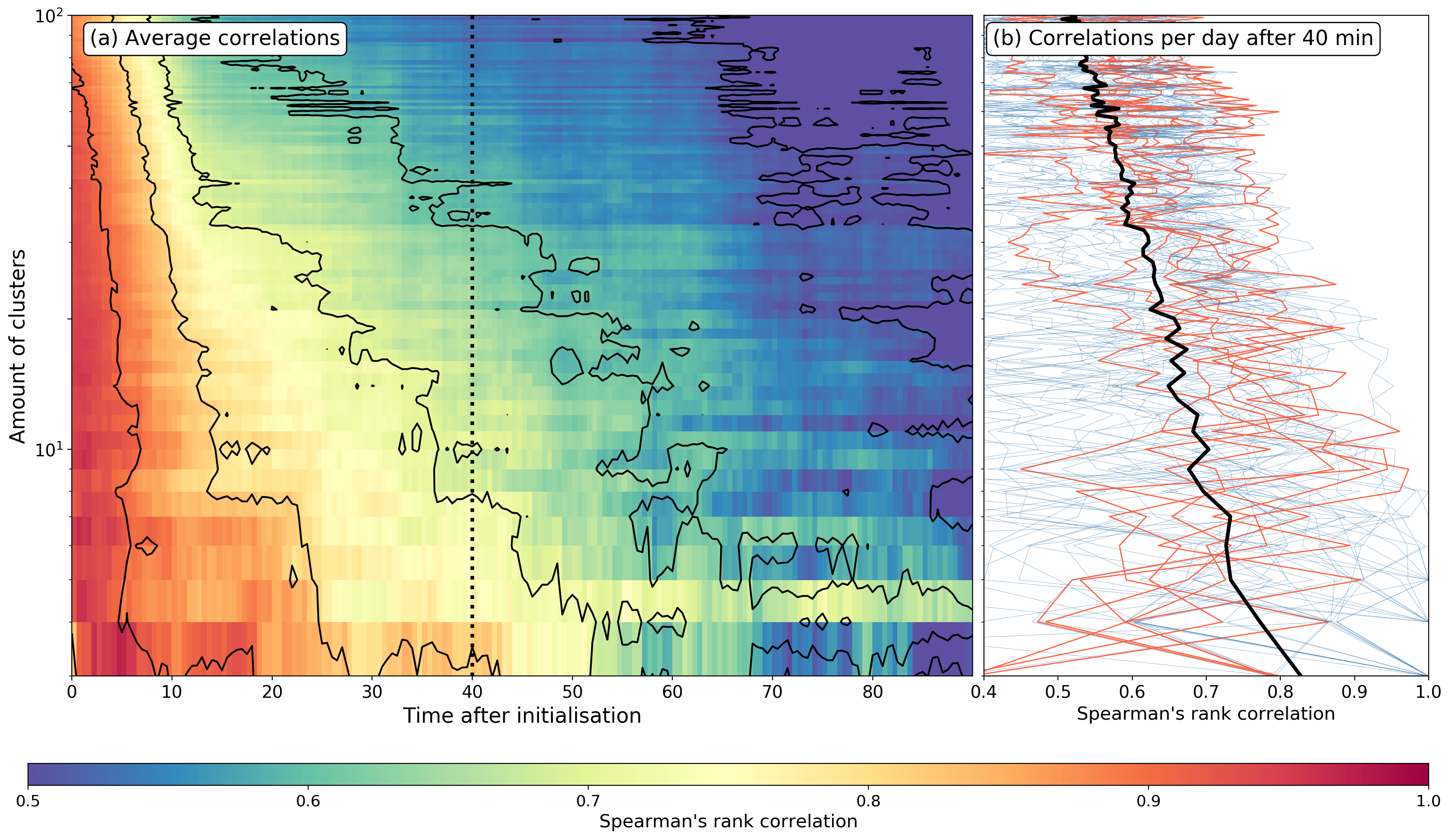}
    \caption{\textbf{Panel (a)}: Average Spearman's rank correlation coefficients in colours averaged over the 50 disrupted days, for various values of $K$ (vertical) and time points after model initialisation (horizontal, in minutes). The contours indicate the levels 0.5, 0.6, 0.7, 0.8 and 0.9. The vertical dashed line corresponds to 40 minutes after initialisation, which is used in the other panel. \textbf{Panel (b)}: Spearman's rank correlation coefficients at 40 min after initialisation. Individual days are split into days that have their maximum at values of $K<9$ (in blue) and those that have their optimum performance at values of $K\geq 9$ (in red). The black line indicates the 50-day average.}
    \label{fig:6_performance}
\end{figure*}

We now turn to the overall performance of the model over the 50 disrupted days. On each of these days, we determine the peak in the total delay and simulate the delays up to two hours after this peak. We then compare what really happened throughout these two hours to what we simulated by computing the Spearman's correlation coefficient $\rho$ at each time point (see Eqn.~\ref{eqn:spearman}). We do this for each number of clusters $K$  ($3\leq K \leq 100$). The average correlations per $K$ and $t$ over the 50 disrupted days is shown in Fig.~\ref{fig:6_performance}a. It is clear that in general, the higher $t$, the lower the correlation. This is intuitively correct: the higher the time after initialisation, the more the model will start differing from reality, for example due to new incoming delays in the real data that are not captured by the model or errors the model that grow with time. In the same panel, we see that higher amount of clusters $K$ also decreases the correlation, which is less obvious. On the one hand, information is lost when coarse graining: for lower values of $K$ detailed information on the positioning of the trains is put together into larger clusters, which may reduce their simulation quality. On the other hand, the diffusion-like spreading is presumably more accurate when looking at a larger scale (lower $K$), since on these scales the  discreteness of delays is averaged out in the data, too. Interestingly, panel (a) also shows bands of $K$ values with near-equal correlations: up to $K=8$, the correlations seem to be more or less the same (very high), at least up to $t = 45$ min. The second band of near-equal correlations is between $8\leq K \leq 17$, followed by a more gradual decay of correlations with $K$, but a sharp decrease in those correlations at $K\approx 27$. One reason for these sudden correlation decreases could be a strong rearranging in the clustering at those $K$ values: e.g., in Fig.~\ref{fig:3_clusters}, panels (a) and (b), one can see that for $K=5$, Brussels is at the border of the red cluster, while at $K=10$, it is in fact in the middle of a cluster. Such rearranging can be quite sudden from one value of $K$ to another. In contrast, the slow decrease in correlation \textit{within} those $K$-bands can be related to a slow change in the clustering structure.

Panel (b) in Fig.~\ref{fig:6_performance} shows the correlation $\rho$ 40 minutes after model initialisation, as function of $K$, for each individual disrupted day. Clearly, these curves seem less gradual as the average displayed in panel (a). In fact, changing $K$ by 1 may impact the correlation up to 0.5 in some exceptional cases --- specifically when $K$ is small (which makes sense as the clustering structure changes rapidly around these values). The average is displayed in black, and the gradual decrease with $K$ is visible. Red, thin lines indicate days on which the maximum correlation is at values $K\geq 9$, which is counteracting the overall pattern we see that the correlation keeps increasing with decreasing $K$ --- on these seven days, the model performs best on a non-trivial level of $K$, between 9 and 11. Appendix~C discusses this non-trivial level (i.e., $K=10$) more in-depth and potential reasons why this clustering configuration may be optimal in some circumstances. The blue lines indicate all other days (where optimal correlations are found with very low values of $K$).

\subsection{Model discussion}

Our results show that modelling delay as a diffusion-like spreading phenomenon clearly has limitations: on the scale of individual stations (Fig.~\ref{fig:4_examples}a), the discrete nature of delays, accompanied with periodic undulations and sharp changes in delay is not simulated at all. Also, delays bound to trains usually travel in a specified direction, which is not captured by our model (which is a weighted form of unidirectional spreading). The diffusion-like-spreading assumption corresponds to the view that the delay propagation is based not on single trains carrying a delay, but on many tiny trains, all going in various possible directions with their portion of the delay, which are randomly chosen, weighted by the variables in the model (e.g., frequency and travel times). Thinking about the diffusion-like model in this way motivates the use of coarse graining to improve the model. Qualitatively we show this in Fig.~\ref{fig:4_examples}b and c and quantitatively this is discussed in Fig.~\ref{fig:6_performance}: a clear increase in performance is visible when comparing results from the clustered version of the model to clustered data. Still, there is a loss of correlation with simulation time: the further from the initialisation time, the more new trains and delays enter the system and change the real delay time series, unaccounted for by the model. It is important to note that the latter will always be a caveat for delay propagation models due to the inherent stochasticity of delay generation.

The toy systems we tested are meant as illustrations to indicate what the model benefits from: average patterns. As soon as the network is dense and there are many trains travelling on it, the real delays spread roughly along simple statistics like train frequencies, which are the basis of this model. But as soon as the network becomes more sparse, especially when it becomes tree-like, the correlation drops.

Our discussion on the classification of disrupted situations showed no clear differences in model performance based on the initial condition, at least as far as its localization (which is the basis of the classification) is concerned. However, our approach was naive: we classified the disrupted days by eye into four groups. We cannot conclude that there are no other, better metrics that do distinguish situations in which the delays spread in a more diffusion-like manner than in other situations --- something we do find in the toy examples. Hence, we propose to investigate such metrics further in future work.

The high performance at low values of $K$ implies that a coarse resolution is better suited for these type of models. The disadvantage of that is the loss of detail. Also, as shown in the toy examples, there are cases we can think of that are not suitable to be modelled well by the model: high sparsity of trains increases the discrete nature of delays and decreases the applicability of the mean-field approximation. Another example where these models have low accuracy is when the delays are mainly governed by stochasticity, and not by propagation dynamics. This is the case in situations where the overall delay level in the network is low. Such situations are difficult to capture well in many delay propagation models, in fact. 

We propose therefore that the model presented here finds its niche in the problem of simulating the propagation of severe delays on a large scale. In such circumstances, the exact magnitude of delays at fixed positions is not always of most interest, while the general trend, speed of delay decay and direction of the overall bulk of delay are of high importance. Such information is well retrievable from the clustered model. In fact, this model is arguably very suited to analyse these large-scale dynamics and how they depend on network topology and high-level parameters such as train frequencies. All information of the system's dynamics is embedded in the $\mathbf{G}$-matrix (Eqn.~\ref{eq:GH}) --- a single matrix that can be analysed using spectral methods to investigate its eigenproperties, for example. Another advantage of this model is its simplicity. Using only a small set of parameters that are easily retrievable from the schedules (which can usually be found online for any European railway system), one can model the whole railways with a single simple differential equation (Eqn.~\ref{eq:difeq}).

\section{Conclusions}
\label{sec:6_Conclusion}

In summary, we devised a model that simulates delays as a diffusion-like spreading phenomenon. The intuition is that, on average, the direction and dissipation of delay relates to aspects of the schedule such as train frequencies and travel times. We apply the model to the Belgian railways and investigate its strengths and weaknesses. In particular, we find that the model performance increases sharply when coarse graining by grouping stations together into clusters. We conclude that this model is mainly of use when working on larger scales and aiming to identify system properties related to delay dissipation and general directions of delay propagation, rather than accurate individual prediction.

We have illustrated the workings and performance of the model using the Belgian railway network as an example. Our framework, however, is general and could be applied to any country. Nevertheless, there are some international differences that should be taken into account. One aspect which may influence the performance of the model in different countries, is the position of important railway hubs. Hubs, such as large cities, are expected to play an important role in the delay propagation dynamics. In Belgium, these hubs are well distributed across the country, apart from the more rural areas in the south-east and far-west. This means that, in our geographic clustering, the hubs usually fall into different clusters. The Netherlands, in contrast, has its most important cities concentrated in the west of the country. In a model with a small number of clusters, it is possible that many of these hubs end up in the same cluster, which may have an unwanted effect on the model's performance and usefulness. For this reason, it might be important to consider other clustering methods. 

Not only geographic differences should be considered when trying to extrapolate these results internationally. Various factors impacting delay propagation vary from country to country, like policy, protocols, infrastructure details and delay statistics in general \cite{Schipper2018}. However, we argue that the increase in model performance when coarse graining is robust to these changes, because the reason for it is of a more theoretical nature: diffusion-like spreading captures average delay fluxes, which are more prominent in clustered systems. 

Applying the model to other countries is straightforward, since its ingredients are general and easily obtained: train turnover rates, frequencies, travel times and adjacency matrices are readily derived from network architecture and railway schedules. A natural extension of our work would thus be to compare the network models for different countries and explore the properties in different spatial resolutions. 

All of the model's dynamics are essentially derived from the matrix $\mathbf{G}$, also for the clustered versions. Exploring the spectral and topological properties of the weighted network that $\mathbf{G}$ describes and relating those properties to the dynamics of the railway system are of interest in our future work. It is possible that a few simple metrics, derived from this matrix, could be used for a quick international comparison of railway networks.

Despite its simplicity, this model already gives us a tool to better understand railway network dynamics. However, there are a number improvements to be made in future work. For example, it is possible to add noise to the model, to account for the generation of new delays. The magnitude and distribution of stochastically generated delays at different stations can possibly be derived from the data. It would be interesting to see how noise may lead to an `equilibrium' delay distribution, in contrast to the highly disrupted situations we considered in this paper. 

Once delay is generated in a real railway system, there are mechanisms and feedback loops that can amplify it or mitigate it. Such mechanisms are not present in the model, but would be a valuable addition. Such feedbacks can be nonlinear, complicating the model but possibly generating new dynamics. Finally, our model is `first-order': the spread of delay is determined only by where it is, not where it came from. Including a memory mechanism will probably increase the accuracy of the model, but may also make it more complex. One way to do this is to not consider delay on stations, but instead on edges. In such a version of the model, the directional information of trains is partially kept, contrary to in our station-based model. 

Diffusion-like spreading is researched in many other fields other than transport literature. In particular, it is well established that the vulnerability versus perturbations of networked systems including social, epidemiological and engineering systems depends on (any quantification of) the modularity \cite{Newman2006}: a more modular dynamical structure prevents large-scale spreading \cite{Balcan2009, dekker2021plos}. Determining the modularity from the $\mathbf{G}$-matrix can be an interesting next step, as the $\mathbf{G}$-matrix does not only incorporate topology, but weights each edge by features from the schedules.

The results in this paper are not only of interest for modellers, but also for railway practitioners. First, the model output can provide insights into system-wide properties, like delay decay and general directions of delay propagation. Second, it is easy to use and all information is embedded in the $\mathbf{G}$ matrix. For example, practitioners might be interested in how isolated regions are from each other: the off-diagonal elements of the $\mathbf{G}$ matrix at the appropriate level of coarse graining reflects how strongly regions are connected, i.e. how much delay flows from one region to another. From an operational point of view, optimal levels of clustering like those seen for the red curves in Fig.~\ref{fig:6_performance}b (see also Appendix~C) can be used to categorise situations, issue protocols or form threat assessments in terms of delays.

We hope that the model itself and the results of the application to Belgium motivates researchers and practitioners to vary the spatial aggregation level to non-trivial levels. We believe these diffusion-like models can offer useful insights on how aspects such as network structure, basic schedule parameters and spatial resolution affect the delay propagation through a railway network. Ultimately, such models can lead to a better understanding of railway delay dynamics.

\begin{backmatter}

\section*{Availability of data and materials}
The Belgian railway delay data for the period January 2019 to May 2020 is available under XYZ. The code for the model and associated $\mathbf{G}$ matrix is available under XYZ. Both will be available upon publication.

\section*{Competing interests}
The authors declare that they have no competing interests.

\section*{Author's contributions}
All authors designed research and investigated the topic. The whole work was coordinated by MMD. ANM acquired, processed and managed data. All authors discussed the theoretical part of the model. Formal derivation of the model was made by JR and solved and formally analyzed by LT, JR and GS. ANM and JR wrote the model implementation and then the whole team worked on the simulations. MMD wrote first draft of the manuscript and all authors reviewed and edited the final text.

\section*{Acknowledgements}
The authors gratefully thank the organisers of the Winter Workshop on Complex Systems 2019, where this project started. They also thank Debabrata Panja for his remarks on the manuscript and thank Matthew Garrod and Maria Waldl for their input at the start of this project. This work is part of the research programme `Improving the resilience of railway systems' with project number 439.16.111. MMD was supported by this research project, which is financed by the Dutch Research Council (NWO) and co-financed by Nederlanse Spoorwegen (NS) and ProRail. JR was supported by an individual fellowship from the Research Foundation Flanders (FWO). ANM was supported by F.R.S-FNRS grant PDR T.0065.19 Collective Footprints and the grant 19-01-00682 of the Russian Foundation for Basic Research. 

\bibliographystyle{bmc-mathphys}
\bibliography{Literature}


\section*{Supplementary Information}
\subsection*{Appendix A --- Model derivations}
Contains important model considerations, like the definition of delays, the derivations of the terms $F_{1,i}$ and $F_{2,i}$ in Eqn.~\ref{eq:Di} and analytical solutions of the model.

\subsection*{Appendix B --- Disrupted days}
Contains a list of the 50 disrupted days and plots of the geographic distribution of delays at the moment of peak delay.

\subsection*{Appendix C --- Non-trivial optimal level of $K$}
Discusses the case of $K=9$, which turns out to be an optimal level of spatial aggregation for a subset of disrupted days.

\subsection*{Appendix D --- Toy models}
Describes the algorithm used to generate toy examples in Fig.~\ref{fig:5_scenarios} and provides example topologies.

\end{backmatter}
\end{document}


\begin{frontmatter}

\begin{fmbox}
\dochead{Research}

\title{Modelling railway delay propagation as diffusion-like spreading --- Supplementary Information}

\author[
   addressref={},                   
   corref={},                       
   email={m.m.dekker@uu.nl}   
]{\inits{MMD}\fnm{Mark M.} \snm{Dekker}}
\author[
   addressref={},
]{\inits{AM}\fnm{Alexey} \snm{Medvedev}}
\author[
   addressref={},
]{\inits{JR}\fnm{Jan} \snm{Rombouts}}
\author[
   addressref={},
]{\inits{GS}\fnm{Grzegorz} \snm{Siudem}}
\author[
   addressref={},
]{\inits{LT}\fnm{Liubov} \snm{Tupikina}}

\end{fmbox}

\end{frontmatter}

\appendix
\renewcommand{\theequation}{\thesection.\arabic{equation}}

\section{Model derivations}
\label{app:model_cons}

In this section we explain how we arrive at an equation for the evolution of the time delays $D_i$, defined on the stations of the network. Recall that the evolution over a time $\Delta t$ is written in general as 
\[ D_i(t+\Delta t) - D_i(t) = F_{1,i}(\vec D) - F_{2,i}(\vec D), \]
where $\vec D$ is the vector with all delays in the network. The two terms above describe how the delay on a station can increase and decrease respectively. We will describe how we obtain an explicit formula for these terms below.

\subsection{Expressing the delay of a train as function of the delay on a station}

Since we aggregate delay on stations but have data on train flows, we need a way of expressing a train's delay $d$ as function of the station's delay $D$ it is contributing to. In other words, we need to find a function $f$ such that:
\[ d_T(t) = f(D_{S(T,t)}(t)), \]
where we have introduced the new notation $S(T,t)$, which is the station \emph{towards which} train $T$ is moving at time $t$. By definition, the delay of a station is the sum of the delays of trains moving to that station. Under assumption that all trains currently moving to $i$ \emph{equally contribute} to the delay $D_i$, this leads to:
\begin{equation}
\label{eq:dT}
d_T(t) = \frac{D_{S(T,t)}(t)}{\# \text{ trains moving to } S(T,t) \text{ at time } t}=\frac{D_{S(T,t)}(t)}{\sum_\text{edges $e$ to ${S(T,t)}$} f_e t_e},
\end{equation}
where to compute the number of trains moving towards a station $i$, we use the formula
\begin{equation}
\# \text{ trains moving to } i\ \text{at time } t = \sum_\text{edges $e$ to $i$} f_e t_e=\sum_{\ell\in\mathcal{N}_\text{in}(i)} f_{\ell i} t_{\ell i}.
\end{equation}
Here $e=(\ell, i)$ refer to edges, $f_e$ denotes the frequency of trains on edge $e$ and  $t_e$ average time it takes to travel edge $e$. The product of $f_e$ and $t_e$ is then the average number of trains on this edge. By  $\mathcal{N}_\text{in}(i)$ we denote the in-neighbour of station $i$, i.e., the stations which have edges directed towards $i$ (similarly we denote the out-neighbours by $\mathcal{N}_\text{out}(i)$). For example, if a station has only one edge outgoing towards it with 2 trains per hour and it takes 15 minutes to cross the edge, on average there is 1/2 train on this edge all the time. 

\subsection{First term: New incoming trains}
\label{app:F1}

First we compute $F_{1,i}(\vec D)$, the contribution of new trains on their way to $i$. These are trains that, in the time interval $[t, \Delta t]$, passed a neighbour $\mathcal{N}_\text{in}(i)$ of $i$  and are now moving towards $i$. At time $t$, these train's delay was counting towards one of the in-neighbours of $i$, but at time $t+\Delta t$ it counts towards $D_i$. The total contribution is
\begin{align*}
F_{1,i}(\vec D) &= \sum_{\substack{\text{Trains $T$ that started on a track to $i$} \\ \text{in the time $\tau\in[t,t+\Delta t]$}}} d_T(\tau)= \\
&= \sum_{j\in\mathcal{N}_\text{in}(i)} \sum_{\substack{\text{Trains $T$ to $j$ that hit $j$} \\ \text{and continue to $i$} \\ \text{in $\tau\in[t, t+\Delta t]$}}} d_T(\tau) \\ 
&\stackrel{\text{Eqn. (\ref{eq:dT})}}{=}  \sum_{j\in\mathcal{N}_\text{in}(i)} \sum_{\substack{\text{Trains $T$ to $j$ that hit $j$} \\ \text{and continue to $i$} \\ \text{in  $\tau\in[t, t+\Delta t]$}}} \frac{D_j(t)}{{{\sum_{\ell\in\mathcal{N}_\text{in}(j)} f_{\ell j} t_{\ell j}}}}=\\
&= \sum_{j\in\mathcal{N}_\text{in}(i)} \# \substack{\text{Trains that come into $j$} \\ \text{and continue to $i$}} \frac{D_j(t)}{\sum_{\ell\in\mathcal{N}_\text{in}(j)} f_{\ell j} t_{\ell j}}=\\
&= \sum_{j\in\mathcal{N}_\text{in}(i)} \# \substack{\text{Trains that hit $j$}}\times \substack{\text{Fraction of trains} \\ \text{that come into $j$} \\ \text{and continue to $i$}}\times \frac{D_j(t)}{\sum_{\ell\in\mathcal{N}_\text{in}(j)} f_{\ell j} t_{\ell j}} \tag{$\star$}
\end{align*}

We now make an approximation: we assume that the fraction of trains that continue from $j$ to $i$ can be computed as follows:
\[ \substack{\text{Fraction of trains} \\ \text{that come into $j$} \\ \text{and continue to $i$}} = \frac{f_{ji}}{\displaystyle\sum_{\ell \in\mathcal{N}_\text{out}(j)} f_{j\ell}} \times(1- \substack{\text{Probability that train}\\\text{has end station at $j$}}) =: r_{ji}\times (1-s_j) =: p_{ji}. \]
This is a strong assumption, as also discussed in the main text: here, we assume that trains, when arriving at any station, pick a random direction on which to continue next, and the probability is weighted by the overall frequency of trains on the edges. In reality, where the train comes from matters for where it will go to.
Note that if $f_{ji}=\text{constant}(j)$ independent on $i$, i.e. all tracks starting from $j$ have the same frequency of trains, then 
\begin{equation*}
    \frac{f_{ji}}{\displaystyle\sum_{\ell \in\mathcal{N}_\text{out}(j)} f_{j\ell}}
\end{equation*}
 reduces to $\left(k_{j,\text{out}}\right)^{-1}$, the out-degree of station $j$. 
 
We also approximate the number of trains that reach station $j$ in the time interval $[t, t+\Delta t$]:
\begin{equation*}
     \# \substack{\text{Trains that hit $j$}} =
     \Delta t \times \sum_{\ell\in\mathcal{N}_\text{in}(j)} f_{\ell j}.
\end{equation*}
In other words, we go over each edge towards $j$ and assume that the number of trains on each edge $e$ that reaches $j$ in a time interval $\Delta t$ is given by $f_e \Delta t$. For example, if there are 6 trains per hour on a given edge, we expect that in an interval of 20 minutes, 2 trains reach the end of the edge. 
 
So now we have
\begin{align*}
\star &= \sum_{j\in\mathcal{N}_\text{in}(i)} \left(\Delta t\sum_{\ell\in\mathcal{N}_\text{in}(j)} f_{\ell j} \right)\times p_{ji}\times \left(\frac{D_j(t)}{\sum_{\ell\in\mathcal{N}_\text{in}(j)} f_{\ell j} t_{\ell j}} \right)= \Delta t\sum_{j\in\mathcal{N}_\text{in}(i)}  p_{ji} B_j D_j(t),
\end{align*}
where we set
\begin{equation*}
    B_j := \frac{\sum_\text{edges $e$ to $j$} f_e}{\sum_\text{edges $e$ to $j$} f_e t_e}=\frac{\sum_{\ell\in\mathcal{N}_\text{in}(j)} f_{\ell j}}{\sum_{\ell\in\mathcal{N}_\text{in}(j)} f_{\ell j} t_{\ell j}}.
\end{equation*}  

\subsection{Term 2: Removal of trains}
\label{app:F2}

The delay of trains that arrive to station $i$ in the interval $[t, t+\Delta t$] has to be subtracted from $D_i$ (since now this train's delay is not counting towards $D_i$, but towards the next station on this train's route). We have
\begin{align*}
F_{2, i}(\vec D)  &= \sum_{\substack{\text{Trains $T$ that reached $i$} \\ \text{in the time $\tau\in[t,t+\Delta t]$}}} d_T(\tau)\stackrel{\text{Eqn. (\ref{eq:dT})}}{=}
 \frac{D_i(t)}{\sum_{\ell\in\mathcal{N}_\text{in}(i)} f_{\ell i} t_{\ell i}} \times \# \substack{\text{trains that reach $i$} \\ \text{in interval $[t, t+\Delta t]$}}.
\end{align*}
The number of trains that arrive to $i$ in the interval $[t, t+\Delta t]$ is
\begin{equation*}
     \sum_\text{edges $e$ to $i$} f_e \Delta t = \Delta t \sum_{\ell\in\mathcal{N}_\text{in}(i)} f_{\ell i},
\end{equation*}
such that we now can state that
\begin{align*}
    F_{2, i}(\vec D)=  \frac{\sum_{\ell\in\mathcal{N}_\text{in}(i)} f_{\ell i}}{\sum_{\ell\in\mathcal{N}_\text{in}(i)} f_{\ell i} t_{\ell i}}D_i(t)\Delta t\stackrel{\text{Eqn. (5)}}{=}\Delta t B_i D_i(t).
\end{align*}

\clearpage
\newpage

\section{Disrupted days}
\label{app:B}

We apply our model to highly disrupted day. They were identified as follows: for each day between January 2019 and May 2020, we calculated the peak total amount of delay in the network. We then took the 50 most delayed days. Most days are from 2019: Jan 15th, Jan 18th, Jan 21st, Jan, 22nd, Jan 23rd, Jan 30th, Jan 31st, Feb 4th, Feb 25th, Mar 10th, Apr 2nd, Apr 29th, May 2nd, May 3rd, May 14th, May 16th, May 27th, Jun 5th, Jun 19th, Jun 24th, Jun 25th, Jun 27th, Jul 23rd, Jul 24th, Jul 25th, Oct 2nd, Oct 6th, Oct 15th, Oct 28th, Nov 4th, Nov 8th, Nov 13th, Nov 14th, Nov 18th, Nov 19th, Nov 20th, Nov 21st, Nov 26th, Nov 28th, Dec 2nd, Dec 4th, Dec 5th, Dec 11st and Dec 27th. Some of them were from 2020: Feb 9th, Feb 10th, Feb 15th, Feb 25th, Mar 17th, May 4th. 

\begin{figure*}
    \centering
    \includegraphics{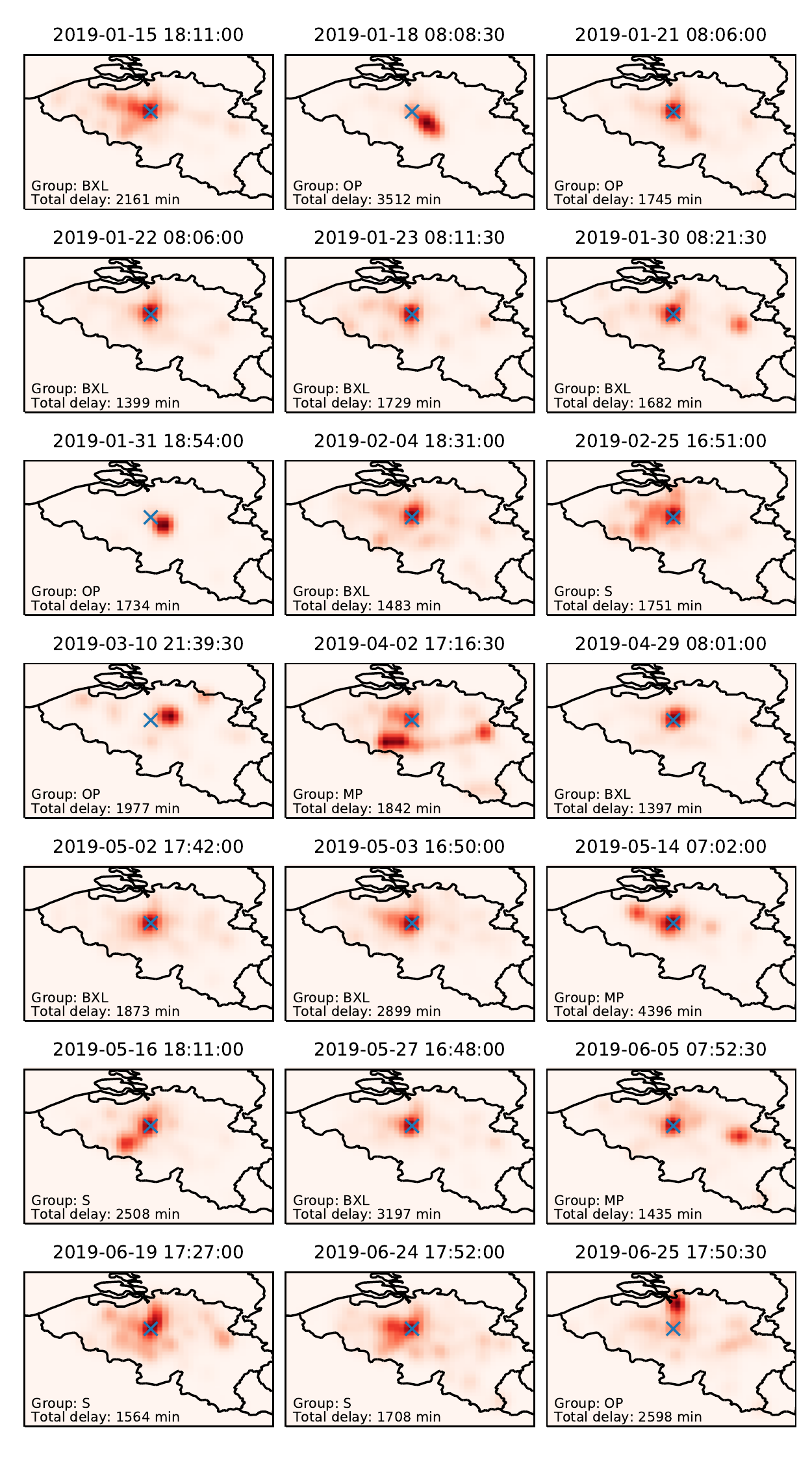}
    \caption{Spatial distribution of time delays on disrupted days.}
    \label{fig:disrupteddays_1}
\end{figure*}

\begin{figure}
    \centering
    \includegraphics{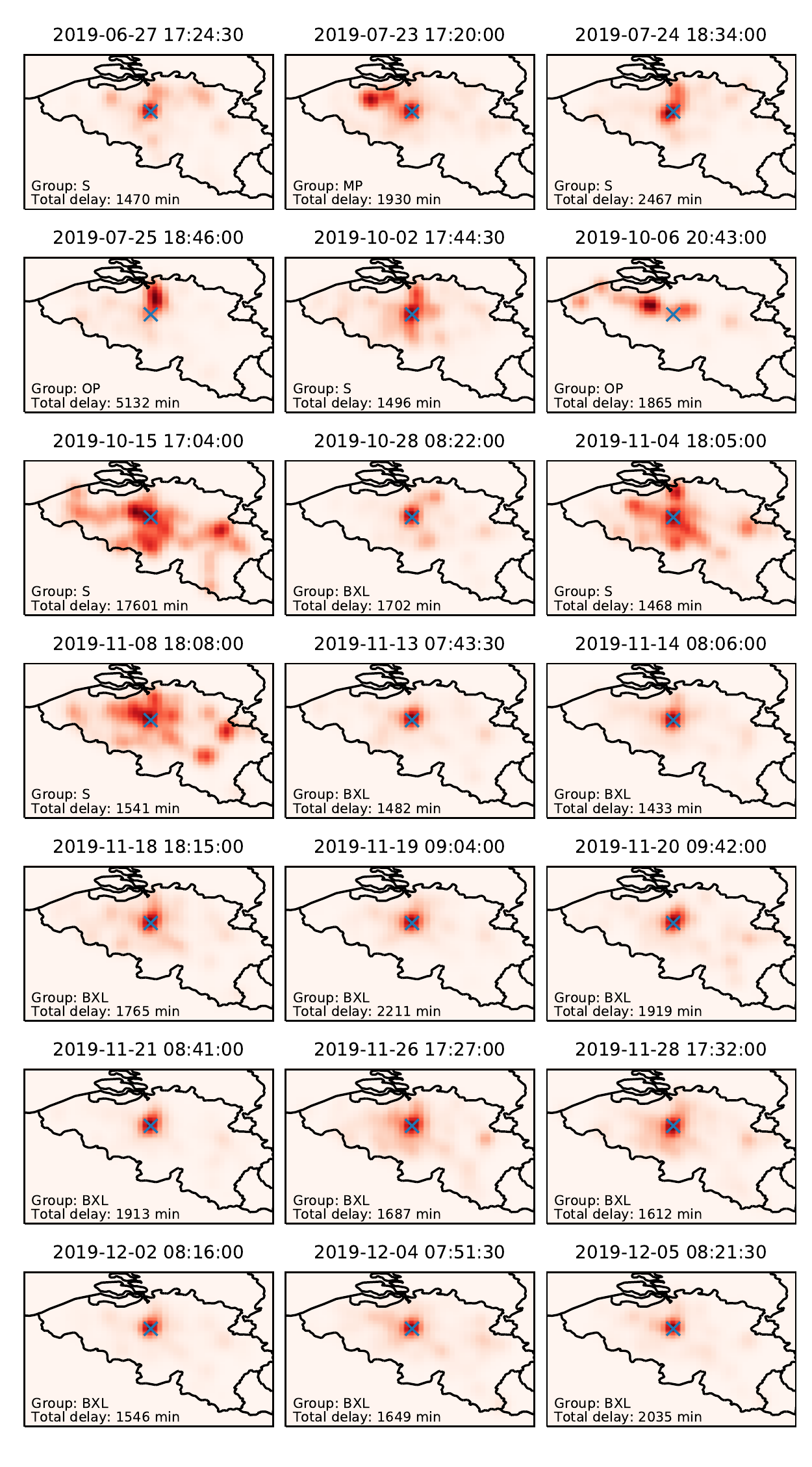}
    \caption{Spatial distribution of time delays on disrupted days.}
    \label{fig:disrupteddays_2}
\end{figure}

\begin{figure}
    \centering
    \includegraphics{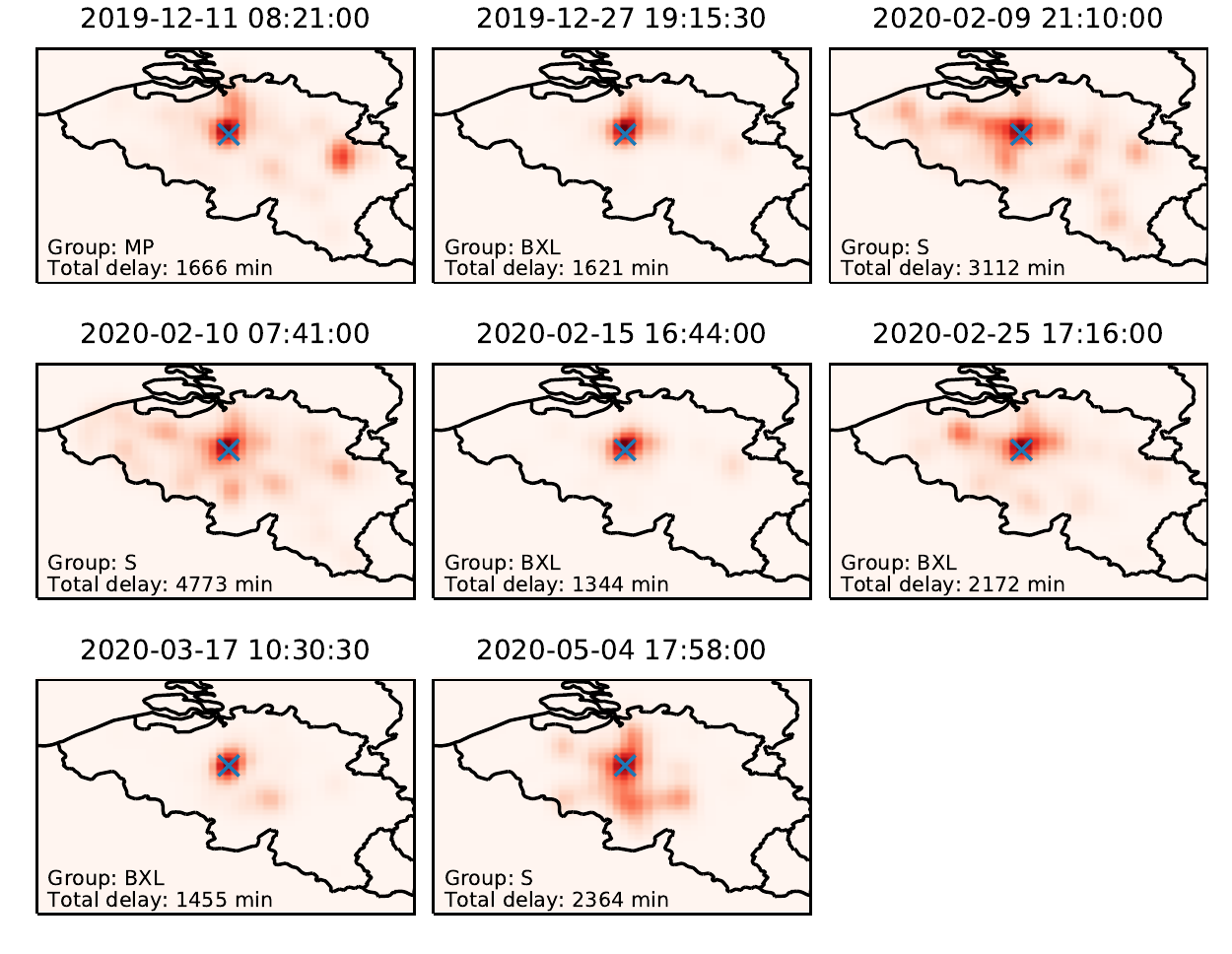}
    \caption{Spatial distribution of time delays on disrupted days.}
    \label{fig:disrupteddays_3}
\end{figure}

Figs.~\ref{fig:disrupteddays_1}, \ref{fig:disrupteddays_2} and \ref{fig:disrupteddays_3} show the situation for these highly disrupted days, at the moment of peak delay. These images were obtained as follows. From the data, we have a list $(s, d)$ of stations and the associated delay at the peak moment. We also have the longitude $x$ and latitude $y$ of each station. We assume that the dataset consisting of $(x, y, d)$ is a sample of a geographical distribution with weights given by the delay. We then use kernel density estimation with a Gaussian kernel with bandwidth 0.1 to obtain a smooth approximation to this distribution. We use the class KernelDensity from the Python package scikit-learn for this. Note that the weights must be positive. We therefore set all delays that are negative (when trains are ahead of schedule) or zero to a small value (0.001). Note that the color on these maps does not hold absolute quantitative information. It only shows the relative magnitude of the delay in different locations. Since the colors are normalized for each day separately, colors between different days cannot be directly compared. 

The maps indicate the time of peak delay on top (up to 30 second resolution) and the total delay in the network at the bottom. The cross marks the location of the station Brussel-Centraal. Each map also mentions in which group we classified the disrupted situation. BXL means that high delays only appear around Brussels, the capital and central hub of the railway network. OP means that there is a single localized peak of delay at another location than Brussels. Situations classified as MP have multiple distinct peaks and finally S stands for `spread out': in these situations the delay is present in large regions.

\clearpage
\newpage

\section{Non-trivial optimal level of $K$}
\label{app:C}

Figure 6 in the main text indicates that on a subset of the 50 disrupted days, performance of the model is optimal for values of $K$ around 10. In this section we discuss possible explanations for this. From a theoretical perspective, we expect the model performance to increase with decreasing $K$, as on a more coarse-grained level, the mean-field approximations in the model are more valid. For example, delay propagation in all possible directions weighted by schedule-based parameters like train frequencies mainly works when there are a lot of trains going in many different directions --- the train frequency then becomes a good weighting factor of where delay really goes. The increasing performance with decreasing $K$ well visible in Fig.~6.

\begin{figure*}[!h]
    \centering
    \includegraphics[width=0.96\linewidth]{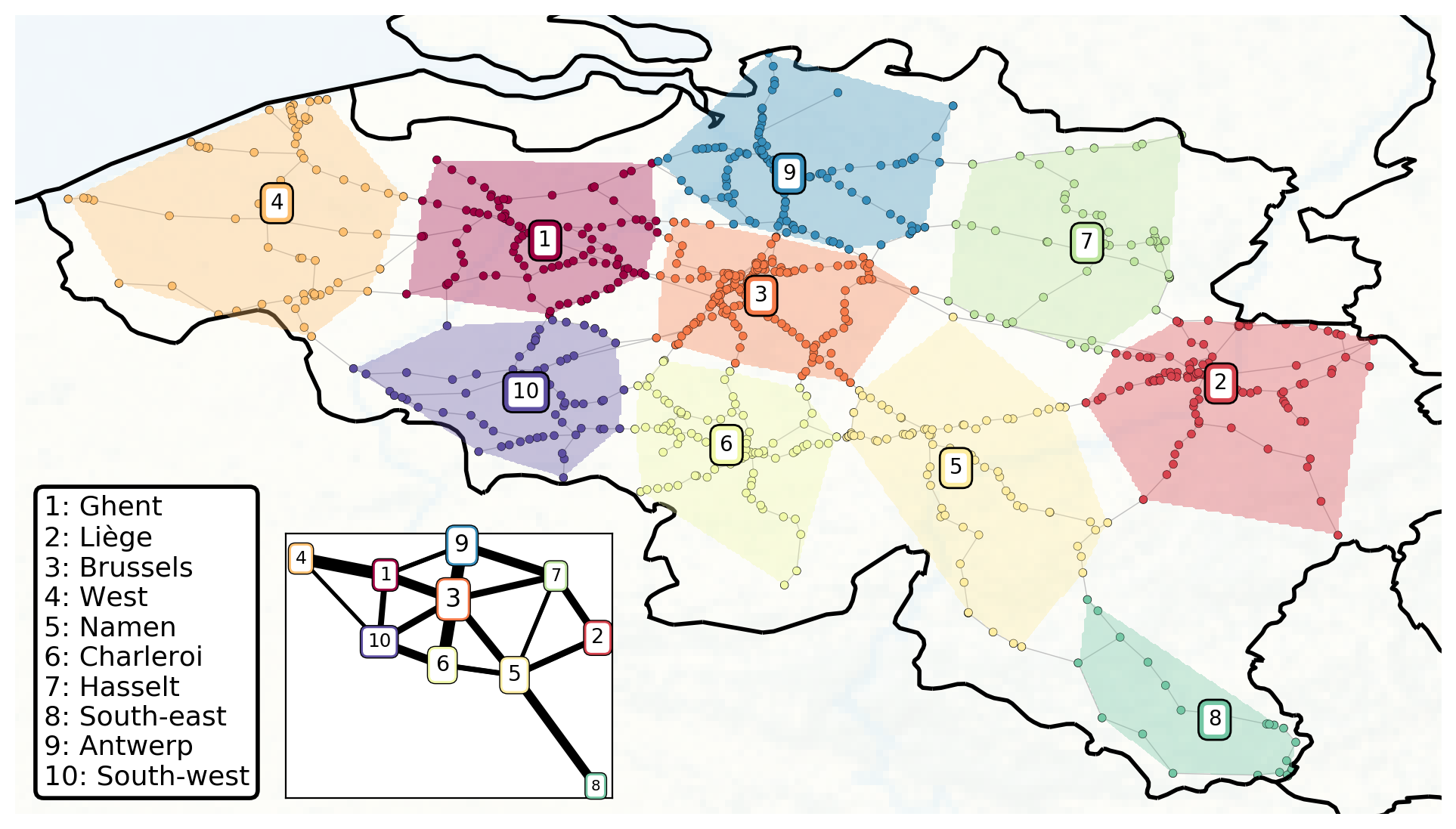}
    \caption{Clusters at optimal aggregation level of 10 clusters. Indicative names of the clusters are shown in a legend. The elements of the resulting 10 by 10 $G$-matrix are reflected in the subpanel at the bottom left: the edge widths show the off-diagonal elements (i.e., dynamical `connectedness' between clusters), node size indicates diagonal elements (i.e., smaller nodes mean stronger delay sinks).}
    \label{fig:a1}
\end{figure*}

Still, on some days, the performance drops when $K$ becomes too high. To investigate possible explanations, we show the clustering configuration at $K = 10$ in Fig.~\ref{fig:a1}. The clusters are approximately equal in size and well distributed across the country. The major cities of Brussels, Antwerp, Li\`ege, Ghent and Charleroi are all located in separate clusters (i.e., 3, 9, 2, 1 and 6, respectively). These cities are also roughly in the center of each of these clusters. Notably, all of the disrupted situations with an optimal performance around $K=10$ fall in the group of situations with delay localized around Brussels (see Appendix~A). 

One possible explanation for the good performance of this number of clusters is as follows. In Fig.~\ref{fig:a1}, the major Belgian cities are centered in different clusters. Such hubs are main routes of delay spreading in severely disrupted situations, but can exhibit complicated dynamics in the real system, since they bring together many trains, from different directions, can have complicated track layouts and scheduling to ensure connections. In our clustered version of the model, having these big hubs centered or well enclosed within clusters entails that the more complicated delay spreading happens within the cluster, compressed into a single diagonal term in the reduced $G$-matrix. The inter-cluster spread mostly contains straight lines between smaller stations and relatively weak connections (often just a few edges between clusters). So, one factor which may explain the $K=10$ optimum is the clustering configuration.

Another factor is of a more theoretical nature. When $K$ becomes too small and the peak delays are at a boundary of a cluster, inter-cluster delay spreading becomes very important and difficult to capture. For example, in Fig.~3a we see that for $K=5$, Brussels --- a location with most delays on disrupted days --- is situated at the boundary of the red cluster. The exact direction spreading just after the peak delay is of high importance because it determines the delays of the yellow and blue cluster to a large extent. Calculating a Spearman's correlation of only five clusters will make it prone to a slight deviation from our model's expected delay spread, even leading up to negative Spearman's correlations for the lowest values of $K$, which is visible at the bottom of Fig.~6. Whether these observations are robust will require more case study-specific research.

\clearpage
\newpage

\section{Toy models}
\label{app:D}

\subsection{Toy model generation}

We aim to test the model on smaller and simple networks. For this, we generate `toy models', involving both an synthetic topology of the infrastructure, as well as an synthetic line schedule on top of this network. We use the following procedure to generate such toy models:
\begin{itemize}
    \item we generate the graph $\mathcal{G}$,
    \item for the fixed visualization layout of the graph $\mathcal{G}$ we determine the Euclidean coordinates of each of the vertices for this layout - these are the geographic coordinates of the station in our example,
    \item we draw $P$ pairs of graph's vertices - they will be the beginnings and the ends of the lines (note: we take each line to be bidirectional in terms of placing trains),
    \item at the beginning of each line we place a train with a random delay, the trains then move at a constant (Euclidean) speed, going towards their destination, along the shortest path in the graph,
    \item at each moment of time we count the delay of each station and compare it through the Spearman's correlation coefficient with our model predictions, averaging over $200$ runs.
\end{itemize}

\subsection{Toy model examples}

While there are endless possibilities of generating the topology of the infrastructure networks for the above toy models, we focus only on two types, of which examples are shown in Fig.~\ref{fig:toygraphs}:
\begin{itemize}
    \item random graphs with $ 15 $ vertices and $20$ edges. If such drawn graph was not connected, we consider its largest connected component.
    \item star graph with $n+1$ vertices, where the vertices $1,\,\dots,\, n$ are connected only to the vertex $0$.
\end{itemize}

\begin{figure*}[!h]
    \centering
    \includegraphics[width=0.125\linewidth]{PaperOutline/Figures/Figure_D:random_graph.pdf}\quad\quad\includegraphics[width=0.45\linewidth]{PaperOutline/Figures/Figure_D:star_graph.pdf}
    \caption{Considered toy-network topologies: random graph with $15$ nodes and $25$ edges (left) and star graph with $9$ nodes (right).}
    \label{fig:toygraphs}
\end{figure*}

\begin{backmatter}

\end{backmatter}